# STRUCTURAL, MAGNETIC AND ELECTRON TRANSPORT PROPERTIES OF ORDERED-DISORDERED PEROVSKITE COBALTITES

*Asish K. Kundu[1] and B. Raveau[2]*

[1]Indian Institute of Information Technology Design & Manufacturing, Dumna Airport Road, Jabalpur –482005, India
[2]Laboratoire CRISMAT, ENSICAEN UMR6508, 6 Boulevard Maréchal Juin, Cedex 4, Caen-14050, France

## ABSTRACT

Rare earth perovskite cobaltites are increasingly recognized as materials of importance due to rich physics and chemistry in their ordered-disordered structure for the same composition. Apart from colossal magnetoresistance effect, like manganites, the different forms of cobaltites exhibit interesting phenomena including spin, charge and orbital ordering, electronic phase separation, insulator-metal transition, large thermoelectric power at low temperature. Moreover, the cobaltites which display colossal magnetoresistance effect could be used as read heads in magnetic data storage and also in other applications depending upon their particular properties. The A-site ordered-disordered cobaltites exhibit ferromagnetism and metal-insulator transitions as well as other properties depending on the composition, size of *A*-site cations and various external factors such as pressure, temperature, magnetic field etc. Ordered cobaltites, having a *112*-type layered structure, are also reported to have an effectively stronger electron coupling due to layered *A*-site cationic ordering. Most importantly for the present article we focus on *La-Ba-Co-O* based ordered-disordered perovskite phases, which exhibit interesting magnetic and electron transport properties with ferromagnetic transition, $T_C \sim 177K$, and it being the first member of lanthanide series. *Zener* double exchange mechanism considered to be crucial for understanding basic physics of the ferromagnetic-metallic phase, yet does not explain clearly the insulating-type phase. In terms of electron transport the ferromagnetic-metallic or insulating/semiconducting states have been discussed in the present article with different types of hopping model.

# INTRODUCTION

Last few decades, there has been extensive research on *ABO₃*-type perovskite oxides of the general formula $Ln_{1-x}Ak_xBO_3$ (*Ln* = trivalent lanthanide, *Ak* = divalent alkaline earth, *B* = transition metal) [1]. Particularly the perovskite cobaltites were discovered in *1950's* and the magnetic ordering was first reported in *1960's* [2], since then there are lots of interesting phenomena reported in the literature. Some of the novel properties of the perovskite cobaltites are known for some time, especially the crystal structure transformation, the paramagnetic *(PM)* to ferromagnetic *(FM)* transition at *Curie* temperature ($T_C$) and the associated insulator-metal transition ($T_{IM}$) and so on. The discovery of colossal magnetoresistance (*CMR*) in doped manganites has renewed great interest in perovskite oxides since the early *90's* [1]. In *1997*, large value of magnetoresistance was reported for newly discovered ordered cobaltite, *LnBaCo₂O₅.₄ (Ln=Eu, Gd)* known as layered *112*-phases [3]. This leads to enhanced interest because of their potential applications in improving magnetic data storage. Moreover, the perovskite cobaltites have also attracted attention because of their possible applications as oxidation catalysts, gas sensors, solid oxides fuel cells and oxygen separation membranes due to their stability in a wide range of oxygen contents, high oxygen diffusivity and electron conduction [4]. Consequently superconductivity was discovered in hydrated sodium cobaltite phase in *2003*, since then cobaltites have received even more attention [5]. Aside from potential applications, the cobaltites exhibit rich phase diagram spanning a wide range of magnetic properties and phenomena like charge ordering, orbital ordering, spin/cluster-glass behavior, electronic phase separation etc [6-11]. These phenomena represent a combined interaction between the spin, the lattice, the charge and the orbital degrees of freedom, which will provide better understanding of strongly correlated electronic behavior. Such interactions are manifested in single crystal, polycrystalline samples as well as in thin films. The properties of perovskite cobaltites can be tuned either by external factors or by chemical means. In certain critical range of cation doping, at the *A*-site of *ABO₃*-perovskite, the rare earth cobaltites exhibit simultaneous occurrence of ferromagnetism and metallicity, along with *CMR* in the vicinity of $T_C$ or $T_{IM}$ [7].

Some of the perovskite-based cobaltites are known to exhibit electronic inhomogeneities arising from the existence of more than one phase in crystals of nominally monophasic composition. This is understood in terms of electronic phase separation described recently in the literature [1]. Such a phenomenon occurs because of the comparable free energies of the different phases. The phase-separated hole-rich and hole-poor regions give rise to anomalous properties such as weak *FM* moments in an antiferromagnetic regime [1]. A variety of magnetic inhomogeneities manifest themselves in $Ln_{1-x}Ak_xCoO_3$ depending on the various factors such as the average radius of the *A*-site cation and size-disorder as well as external factors such as temperature, magnetic field etc [6-11]. In the last few years electronic phase separation in cobaltites has attracted considerable attention.

In this article, we discuss the *A*-site cationic ordering and disordering effects on magnetic and electron transport properties for rare earth cobaltites. Perovskite cobaltites have two possible forms of the *A*-site cations distribution depending on the type of cations or the synthesis procedures [9, 12]. The first reported compounds on perovskite cobaltites are the *A*-site disordered structure [2, 13], which have been investigated for last few decades, and the other one is *A*-site ordered perovskites possessing a layered *112*-type structure [3]. The latter one discovered few years back, consists of oxide layers *[LnO]-[CoO$_2$]-[AkO]-[CoO$_2$]* alternating along the *c*-axis [3, 9, 12, 14, 15]. The ordering of *Ln$^{3+}$* and *Ak$^{2+}$* ions is favorable if the size difference is large between the *A*-site cations, hence smaller size *Ln$^{3+}$* and bigger *Ak$^{2+}$* ions easily form a layered *112*-structure and till date it is reported only for *Ak$^{2+}$ = Ba$^{2+}$* cation i.e. *LnBaCo$_2$O$_{5+\delta}$ (0 ≤ δ ≤1)* series [3, 9, 12, 14-16]. In contrast, as the size difference becomes smaller, as for example between *La$^{3+}$* and *Ba$^{2+}$*, the disordered cubic perovskite becomes more stable and special treatment is required to obtain the *112* ordered structure, as reported recently for the order-disorder phenomena observed in the perovskites of the system *La-Ba-Co-O* [12, 16]. The *CoO$_2$* layers in ordered cobaltites are free from the random potential which would otherwise arise from the *Coulomb* potential and/or local strain via the random distribution of *A*-site cations (*Ln$^{3+}$/Ak$^{2+}$*). Both ordered and disordered cobaltites exhibit similar features like ferromagnetism, insulator-metal transition and magnetoresistance in a certain temperature range, yet prominent differences are evidenced in their properties. The ferromagnetism in these cobaltites is considered to be due to favorable *Co$^{3+}$ - O - Co$^{4+}$* interactions. The effect of random distribution of *A*-site cations or disordered perovskite structure arising from the chemical disorder as observed in the conventional random doping at the *A*-site in *ABO$_3$*-perovskite. For rare earth manganites, *Attfield et al.* [17] have reported the role of *A*-site cationic radius, $\langle r_A \rangle$, and size disorder parameter, $\sigma^2$, on the $T_C$ and/or $T_{IM}$ with variation of these parameters. In the present article we have also discussed the role of $\langle r_A \rangle$, and $\sigma^2$ for disordered cobaltites *Ln$_{0.5}$Ak$_{0.5}$CoO$_{3+\delta}$* (*Ln = La, Nd, Gd and Ak = Ba, Sr*) which unusually influence the magnetic and electronic properties [13, 18, 19]. These interesting phenomena are related to structural disordering caused by the substitution of *Ak$^{2+}$* ions in place of *Ln$^{3+}$* and we have briefly presented the effect of size disorder $\sigma^2$ for *Ln = Nd, Gd* [18, 19]. The structural disorder in cobaltites is *3D*, while a layered *2D* structure is adopted by the *112*-phase ordered cobaltites, *LnBaCo$_2$O$_{5+\delta}$*. The crystallographic structures for perovskite cobaltites with an integral number of oxygen ions per formula unit are well known, no consensus has been reached for non-integer compounds. Basic knowledge of the crystallographic structure of a compound is of particular importance for determining magnetic structures, since these two properties are closely related. Hence we have presented first the crystal structure description for both the (ordered-disordered) cobaltites before discussing the physical properties.

## CRYSTAL STRUCTURE

In general, the rare earth ordered-disordered cobaltites crystallize in the perovskites structure with various types of superstructures also evidenced. The disordered *ABO$_3$*-perovskite is a simple cubic structure (*Pm-3m*) as shown in Figure 1. However, many perovskite deviate a little from this structure even at room temperature. The perovskite structure is most stable when the *Goldschmidt* tolerance factor, *t*, is unity (for cubic structure), which is defined by, $t = (r_A + r_O)/ \sqrt{2} (r_B + r_O)$ where, $r_A$, $r_B$ and $r_O$ are the average ionic radius of the *A, B* and *O* ions respectively. Deviation of '*t*' from unity leads to the structural distortion. For a small deviation in *t* (i.e. *t* < 1), the crystal structure changes from cubic to tetragonal, rhombohedral or orthorhombic etc. In this situation the ⟨Co-O-Co⟩ bond angle decreases from *180°*. The perovskite structure occurs only within the range *0.75≤ t≤ 1.00*. The stability of the perovskite structure of cobaltites depends on the relative size of the *Ln/Ak* and *Co* ions in $Ln_{1-x}Ak_xCoO_3$. In disordered cobaltites *Ln/Ak* cation is surrounded by eight corner sharing $CoO_6$ octahedra, which build a *3D* network. The smaller ionic radius of the cations results in a lower value of '*t*', consequently increasing the lattice distortion. The increase in lattice distortion significantly decreases the ⟨Co-O-Co⟩ bond angle from *180°*, which strongly affects the physical properties of cobaltites. When *t* < 1, there is a compression of the *Co-O* bonds, which in turn induces a tension on *Ln-O* bonds. A cooperative rotation of the $CoO_6$ octahedra and a distortion of the cubic structure counteract these stresses. The rare earth cobaltites can be crystallized not only in cubic or orthorhombic structure (as reported mostly), but also in tetragonal, hexagonal, rhombohedral, and monoclinic structures as well. As a consequence, for the disordered cobaltite series $Ln_{0.5}Ba_{0.5}CoO_{3+\delta}$ the crystal structure is reported as cubic or rhombohedral for *Ln = La*, whereas for other lanthanides such as *Pr, Nd, Gd, Dy* the systems crystallize in the orthorhombic structure with different space groups [13, 18]. The structural *Rietveld* analysis for disordered $Ln_{0.5}Sr_{0.5}CoO_3$ series reveals that the structure is rhombohedral for *Ln = La, Pr Nd* and that for Gd, the structure is orthorhombic as reported for *Ba*-doped compounds [13].

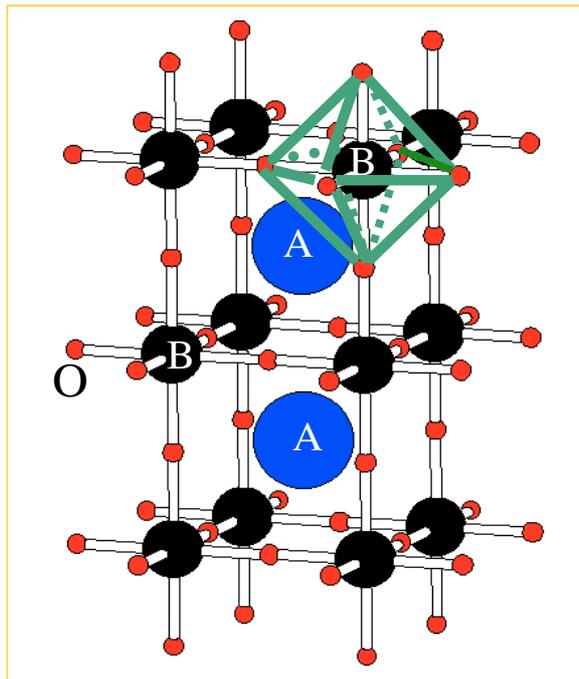

Figure 1. Rare earth $ABO_3$ perovskite with disorder structure.

Similarly, the layered *112*-type ordered cobaltites $LnBaCo_2O_{5+\delta}$ can be found in several crystallographic symmetries at room temperature. The variation of oxygen stoichiometry (*0 ≤ δ ≤1*) in ordered *112*-type cobaltites lead to various structures as well as different cobalt and oxygen coordination's such as pyramidal, octahedral and/or the mixing of both environments for the *Co*-ions [14, 15]. In the following, the room temperature structures for different values of '$\delta$' will be discussed. The crystal structure of the stoichiometric *LnBaCo$_2$O$_5$* (*Ln = Eu, Gd, Tb, Ho*) cobaltite is tetragonal with *P4/mmm* space group (unit cell $a_p \times a_p \times 2a_p$; where $a_p$ is defined as pseudo cubic cell parameters) [14, 15]. This corresponds to a doubling of the original perovskite unit cell along the *c*-direction due to alternating *BaO* and *LnO$_\delta$* layers (Figure 2). The layered structure is best observed in the most oxygen deficient case *LnBaCo$_2$O$_5$*, because it is assumed that the oxygen ions are absent only in the *Ln*-layer [14, 15]. For $\delta = 0$ the $Co^{2+}$ and $Co^{3+}$ ions (ratio 1:1) are all within square base pyramids formed by their five oxygen neighbors.

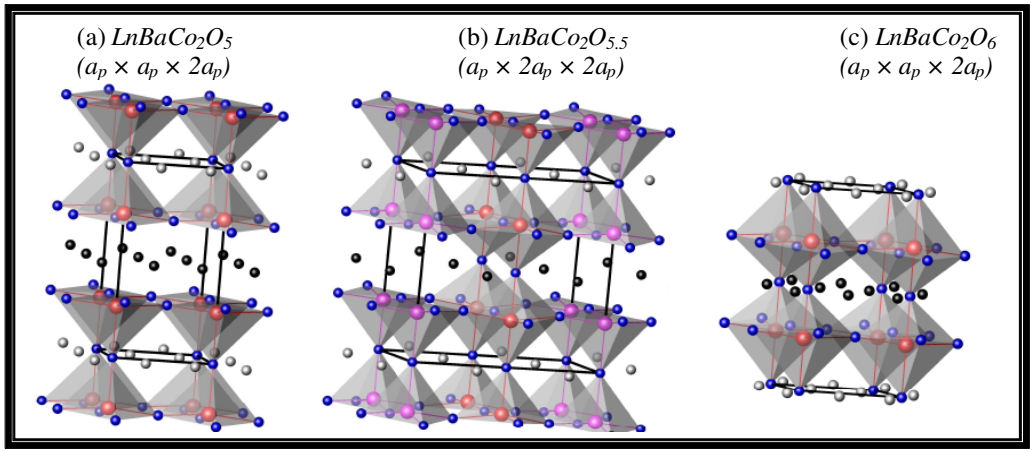

Figure 2. Layered *112*-type ordered *LnBaCo$_2$O$_{5+\delta}$* $(0 \leq \delta \leq 1)$ cobaltites with (a) $\delta=0.0$; *LnBaCo$_2$O$_5$*, (b) $\delta=0.5$; *LnBaCo$_2$O$_{5.5}$* and (c) $\delta=1.0$; *LnBaCo$_2$O$_6$*.

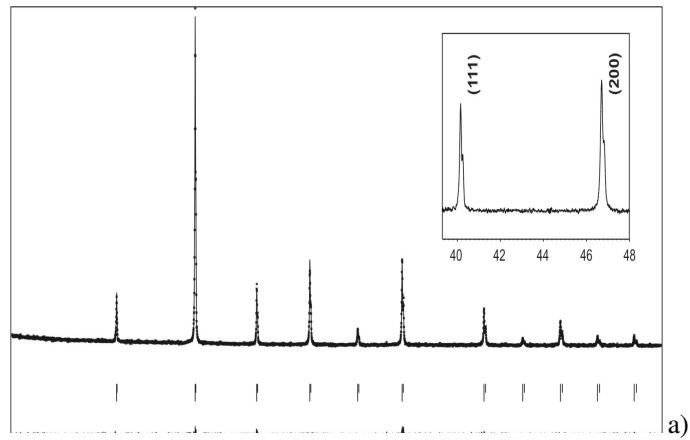

Figure 3. (Continued)

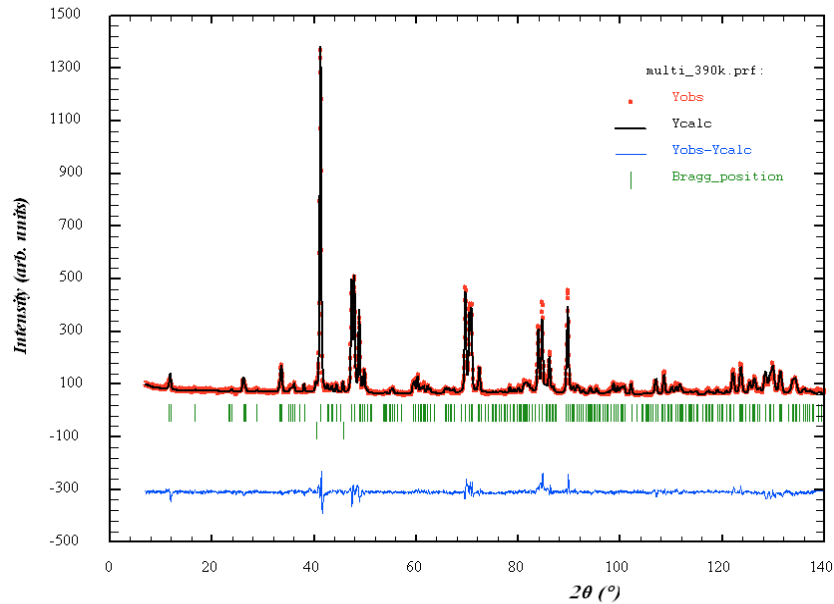

b)

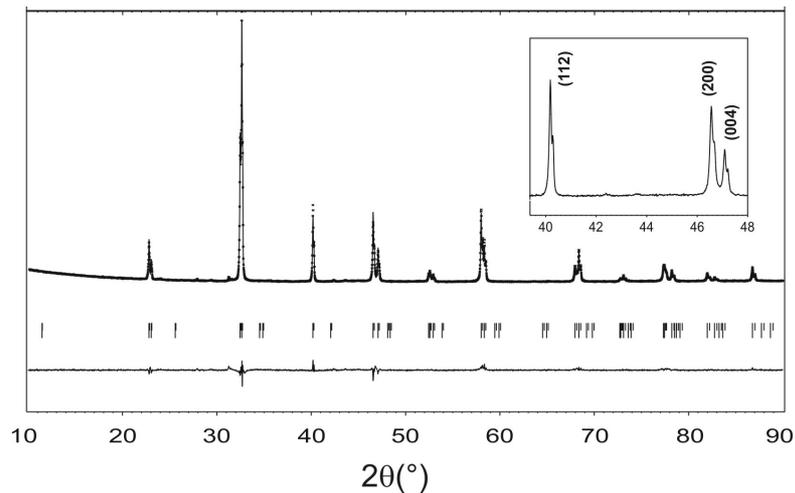

c)

Figure 3. Rietveld refinement of X-ray diffraction pattern for (a) disordered $La_{0.5}Ba_{0.5}CoO_3$ (b) ordered $LaBaCo_2O_{5.5}$ and (c) ordered $LaBaCo_2O_6$ (Taken from Ref. 12 & 16).

For non-stoichiometric cobaltites $LnBaCo_2O_{5+\delta}$ ($0 < \delta < 1$) the crystal structure is more complex: due to oxygen vacancy ordered super-structures can arise, which vary with oxygen content. The oxygen content is strongly dependent on the size of the lanthanides [14]. Clearly, a strong correlation exists between the $Ln^{3+}$ radius and the amount of oxygen the compound can accommodate. The first structural study of an ordered oxygen deficient perovskite of type $LnBaCo_2O_{5+\delta}$ was presented for the series $Ln = Pr\ to\ Ho$ [14]. Although, the series has been investigated in details for almost all lanthanide elements of the periodic table, interestingly there was no such report on the first member of this series, i.e. for $LaBaCo_2O_{5.5}$. Very recently we have reported $LaBaCo_2O_{5.5}$ and characterized by neutron diffraction, electron microscopy and magnetic studies, showing that at room temperature the structure is *112-* layered orthorhombic $a_p \times 2a_p \times 2a_p$ supercell with *Pmmm* space group [16]. In contrast, for *Ln = Pr – Ho*, all of the x-ray diffraction patterns could be indexed using a tetragonal structure $a_p \times a_p \times 2a_p$ with *P4/mmm* symmetry [14]. However, electron diffraction measurements additionally revealed two kinds of superstructures, depending on the radius of the lanthanide (or the oxygen content). For larger lanthanides (*Pr, Nd, Sm, Eu, Gd and Tb*) a doubling of one lattice parameter is observed, corresponding to an orthorhombic *Pmmm* $a_p \times 2a_p \times 2a_p$ supercell, whereas for smaller lanthanides (*Ho and Dy*) a tripling of two lattice parameters is noticed, as in a $3a_p \times 3a_p \times 2a_p$ supercell. It was suggested [14] that ordering of oxygen vacancies is at the origin of the observed superstructures. Importantly, the authors found that the oxygen vacancies are located uniquely in the $LnO_\delta$ layers (apical positions). By changing the oxygen content the superstructures either vanish or change from one to the other: Reducing oxygen content in $GdBaCo_2O_{5+\delta}$ from $\delta = 0.4$ to $0$ destroys the superstructure. On the other hand in $HoBaCo_2O_{5.3}$ an increase of $\delta$ from $0.3$ to $0.4$ leads to a change in superstructure from $3a_p \times 3a_p \times 2a_p$ to $a_p \times 2a_p \times 2a_p$. High resolution electron microscopy supported the assumption that the ordering of the oxygen vacancies is responsible for the superstructures. *Burley et al.* and *Pralong et al.* [14] have reported $NdBaCo_2O_{5+\delta}$ for various oxygen stoichiometries. The $\delta = 0$ compound has a tetragonal structure with $a_p \times a_p \times 2a_p$ supercell where the *Nd* layer incorporates all the oxygen vacancies. For the slightly higher oxygen content $\delta = 0.38$ the structure is derived from the $\delta = 0$ one, but with oxygen ions inserted randomly into the $NdO_\delta$ layers. Oxygen vacancy ordering in *b*-direction resulting in an orthorhombic $a_p \times 2a_p \times 2a_p$ supercell unit cell with *Pmmm* symmetry is reported for $\delta = 0.5$, in agreement with other reported structure [14]. The oxidized material with $\delta = 0.69$ is again described by a tetragonal $a_p \times a_p \times 2a_p$ supercell unit cell, but a very weak peak originating from a doubling of the unit cell along the *b*-direction was detected.

Finally, for stoichiometric cobaltites $LnBaCo_2O_6$ all *Co*-ions ($Co^{3+}$ and $Co^{4+}$; ratio 1:1) are in octahedral environment. In the $LnBaCo_2O_6$ series, the first member *Ln = La* could be synthesized in the two forms, ordered and disordered as shown by neutron diffraction and electron microscopy [9, 12]. At room temperature and in normal synthesis condition the crystal structure is cubic for the disordered cobaltite, which means the $La^{3+}$ and $Ba^{2+}$ ions are distributed randomly on the *A*-site. This is due to the smaller size differences between the cations as discussed earlier. Nevertheless, the ordered *112*-phases $LaBaCo_2O_{5.5}$ and $LaBaCo_2O_6$ are obtained for the same composition in special synthesis conditions [12, 16], which crystallize at room temperature in orthorhombic and tetragonal structures respectively. Therefore, a layered *112*-type ordered perovskite structure $LnBaCo_2O_6$ with a smaller lanthanide also exist, again with a tetragonal unit cell $a_p \times a_p \times 2a_p$ with *P4/mmm* symmetry, but require special synthesis conditions as reported by *Pralong et al.* [14]. To conclude the ordered cobaltites phase, a summary of the most commonly used models for compounds with oxygen content $\delta = 0, 0.5, 6$ at room temperature is given (Figure 2):

1. $LnBaCo_2O_5$ has the tetragonal structure with $a_p \times a_p \times 2a_p$ supercell (*P4/mmm* symmetry). $Ln^{3+}$ layers alternate with *BaO* layers along the *c*-axis (oxygen vacancies in *Ln* layer).
2. $LnBaCo_2O_{5.5}$ has an orthorhombic structure with $a_p \times 2a_p \times 2a_p$ supercell (*Pmmm* symmetry). $LnO_{0.5}$ layers alternate with *BaO* layers along the *c*-axis. Along the *b*-direction the oxygen vacancies are ordered. This leads to an alternation of $CoO_5$ pyramids and $CoO_6$ octahedra along the *b*-direction.

3. $LnBaCo_2O_6$ is again tetragonal unit cell $a_p \times a_p \times 2a_p$ with $P4/mmm$ symmetry. $LnO$ layers alternate with $BaO$ layers along the $c$-axis (no oxygen vacancies in $LnO$ layer).

We will now discuss in brief the various parameters to obtain different phases (ordered-disordered) of $La$-$Ba$-$Co$-$O$ cobaltites. The synthesis of $LaBaCo_2O_{5.5}$, maintaining the "$O_{5.5}$" stoichiometry and a perfect layered ordering of $La^{3+}$ and $Ba^{2+}$ cations, is delicate due to their small size difference which favors their statistical distribution. Moreover, the larger size of $La^{3+}$ compared to other lanthanides allows large amounts of oxygen to be inserted, so that the disordered $La_{0.5}Ba_{0.5}CoO_3$ perovskite is more easily formed under normal conditions. Thus, the successful synthesis of ordered phase requires several steps, using soft chemistry method, and the strategy was to control the order-disorder phenomena in this system by means of two synthesis parameters, temperature and oxygen partial pressure. In order to favour the ordering of the $La^{3+}$ and $Ba^{2+}$ cations, the synthesis temperature was as low as possible, and consequently a soft-chemistry synthesis route was used since it allows a high reactivity at low temperature. However, this condition is not sufficient alone to achieve a perfect ordering of these cations. The formation of $La^{3+}$ layers is in fact favoured by the intermediate creation of ordered oxygen vacancies, leading then to the $112$-type layered non-stoichiometric cobaltites $LaBaCo_2O_{5+\delta}$, built up of layers of $CoO_5$ pyramids between which the $La^{3+}$ smaller than $Ba^{2+}$ cations can be interleaved. In this process, it is rather difficult to control the oxygen stoichiometry to "$O_{5.5}$". For this reason, the synthesis of the ordered $LaBaCo_2O_6$ phase was carried out initially, using high purity argon gas, followed by annealing in an oxygen atmosphere at specific temperature. Thereafter, the layered $112$ cobaltite $LaBaCo_2O_{5.5}$ was obtained from the ordered $LaBaCo_2O_6$ phase by employing temperature controlled oxygen depletion method in inert atmosphere [16].

The X-ray powder diffraction (*XRD*) patterns of the three perovskite phases (Figure 3) show their excellent crystallization. The disordered phase $La_{0.5}Ba_{0.5}CoO_3$, (Figure 3a) and the ordered phase $LaBaCo_2O_6$ (Figure 3c) exhibit sharp peaks, the patterns are indexed with the cubic $Pm$-$3m$ and tetragonal $P4/mmm$ space groups, respectively [12]. The latter corresponds to a doubling of the cell parameter along the $c$-axis related to the 1:1 ordering of the $LaO/BaO$ layers (Figure 2c). More importantly, the $XRD$ pattern of the ordered $LaBaCo_2O_{5.5}$ (Figure 3b) is completely different from that observed for the ordered phase (Figure 3c) and for this reason it is indexed and refined using the orthorhombic structure $Pmmm$ space group [16]. The result for ordered-disordered cobaltites is slightly different from that obtained by *Nakajima et al*. [9], who observed a smaller cell volume for the ordered phase ($58.66$ $Å^3$/$Co$) compared to the disordered one ($58.77$ $Å^3$). Remarkably, the $La/Ba$ ordering also involves a slight deformation of the perovskite sublattice with a dilatation of the $a_p$ parameter within the $LaO/BaO$ layers and a compression along the $LaO/BaO$ layers stacking direction.

Transmission electron microscopy (*TEM*) investigations on the disordered $La_{0.5}Ba_{0.5}CoO_3$ and ordered $LaBaCo_2O_6$ cobaltites confirm their cubic and tetragonal structures respectively. The selected area electron diffraction (*SAED*) patterns of the disordered cobaltite and the corresponding high resolution electron microscopy (*HREM*) images (Figure 4a) are indeed characteristic of a classical cubic perovskite ($Pm$-$3m$) with $a \approx a_p \approx 3.9$ Å. For ordered $LaBaCo_2O_6$, the reconstruction of the reciprocal space from the *SAED* patterns leads to a tetragonal cell, with $a \approx a_p$ and $c \approx 2a_p$, compatible with the space group $P4/mmm$. The doubling of one cell parameter with respect to the simple perovskite cell is clearly observed on the *HREM* image displayed in Figure 4b and on the corresponding *SAED* [*100*] zone axis patterns (inset of Figure 4b). Beside the *SAED* investigation which gives a better view of the microstructure, bright field images are also of great interest to obtain information about the size of the *112*-type domains, taking into consideration the possibility of twinning. In the case of the disordered cubic cobaltite no twinning is observed, as expected from the symmetry. In contrast, in the long-range ordered *112*-perovskite *LaBaCo$_2$O$_6$* twinning is clearly evidenced (Figure 5b).

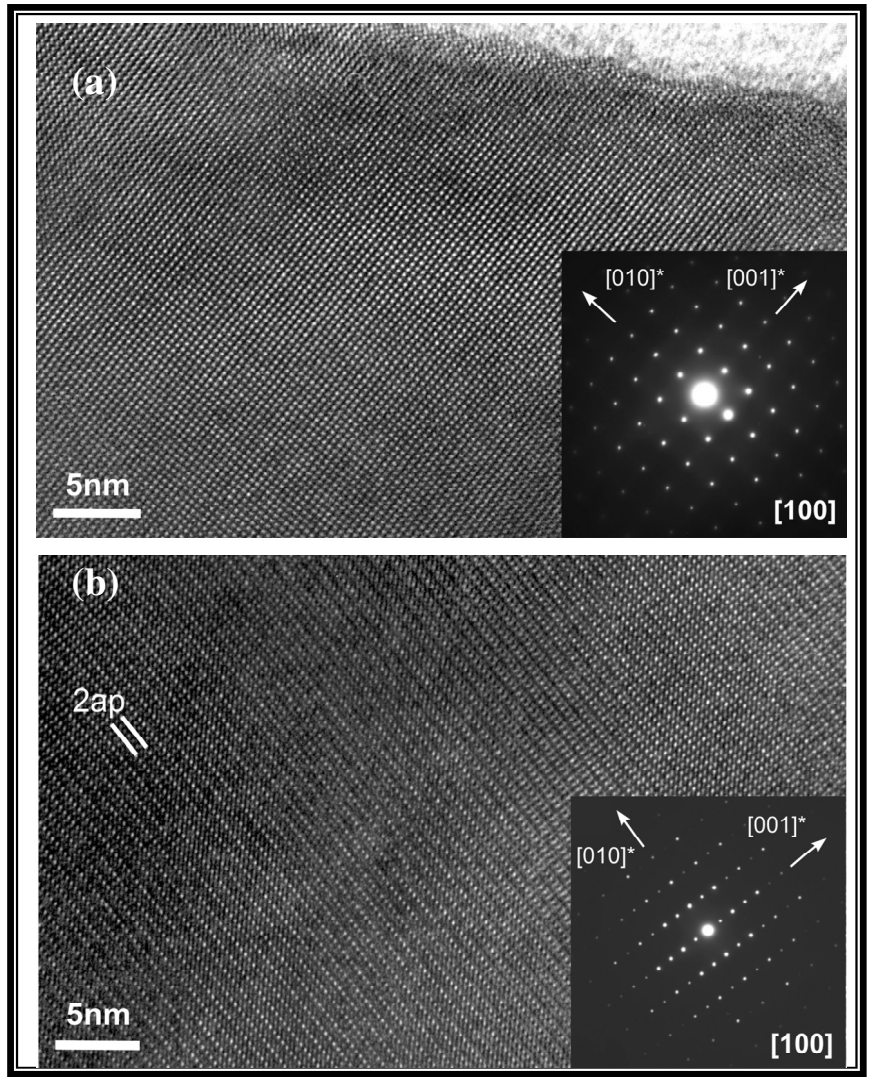

Figure 4. The HREM images and the corresponding SAED patterns for (a) disordered *La$_{0.5}$Ba$_{0.5}$CoO$_3$* and (b) ordered *LaBaCo$_2$O$_6$* (Taken from Ref. 12).

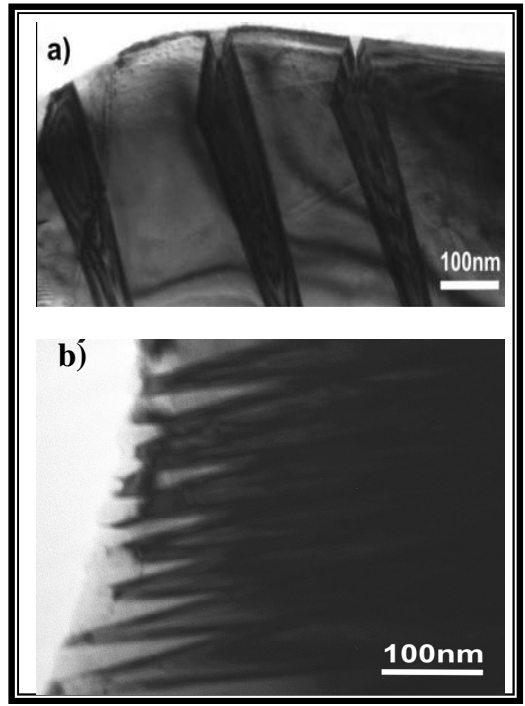

Figure 5. Bright field images for ordered cobaltites (a) $LaBaCo_2O_{5.5}$ and (b) $LaBaCo_2O_6$ (Taken from Ref. 12 & 16).

Moreover, *Rietveld* refinements of the neutron powder diffraction (*NPD*) data of $LaBaCo_2O_{5.5}$, shows a deviation from the ideal orthorhombic *Pmmm* structure due to partial disorder of the oxygen atoms at the $LaO_{0.5}$ layers. One indeed observes occupancy of about *90%* and *10%* of the *O6* and *O7* sites by oxygen, respectively, instead of the expected 100% and 0% [16]. Therefore, an investigation of ordered $LaBaCo_2O_{5.5}$ cobaltite by *TEM*, in order to identify possible secondary phase, superstructures, and/or particular microstructural features is of particular interest. In the Figure 6a, *SAED* confirms that the observed spots can be indexed with the *Pmmm* $a_p \times 2a_p \times 2a_p$ structure refined from powder diffraction. However, two additional features are also evidenced: First, one observes twinned domains at a microscale level, all over the investigated crystals regions, corresponding to the *Pmmm* $a_p \times 2a_p \times 2a_p$ structure *(112-type I)*, second, in few areas of some investigated crystals, very weak extra reflections are evidenced that cannot be indexed considering only the *Pmmm* $a_p \times 2a_p \times 2a_p$ structure. Considering the literature on the *112*-type compounds, a part of these extra spots can be indexed considering the *Cmmm* $2a_p \times 4a_p \times a_p$ cell corresponding to the $LaBaMn_2O_{5.5}$ compound [20]. This is in agreement with the results of the *NPD* refinements where the occupancies obtained for the *O6* and *O7* atomic positions can be locally attributed to a different vacancy/oxygen ordering (Figure 6b) leading notably to the existence of faulted zones having a centered structural motif *(112-type II)*. This tendency is evidenced by *SAED* but also visible in *HREM* for instance looking at one of the directions (Figure 6a). In summary, the $LnBaCo_2O_{5.5}$ structure consists mainly of 112 type I type domains (90%) combined with 112 type II (manganites) domains (10%). Finally, it is worth mentioning that no $3a_p \times 3a_p \times 2a_p$ superstructure observed by TEM for ordered $LaBaCo_2O_{5.5}$ cobaltite, as evidenced for $HoBaCo_2O_{5.5}$ and $YBaCo_2O_{5.44}$ compounds [14, 21].

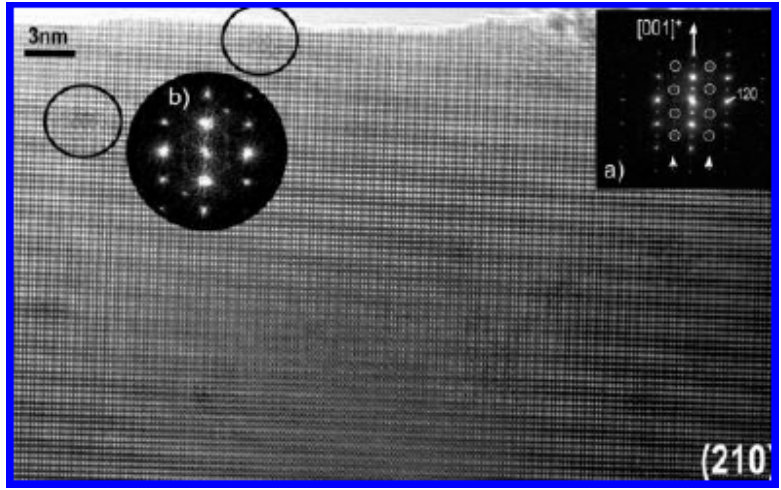

Figure 6a. The HREM images and the corresponding SAED patterns for ordered $LaBaCo_2O_{5.5}$.

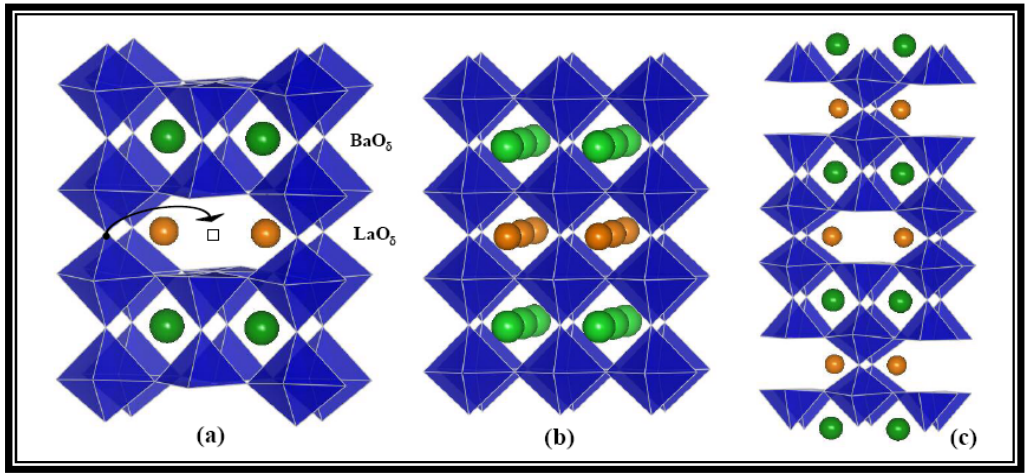

Figure 6b. Layered 112-type ordered $LaBaCo_2O_{5+\delta}$ cobaltites with (a) $\delta=0.5$; $LnBaCo_2O_5$,(type-I) (b) $\delta=1.0$; $LnBaCo_2O_6$ and (c) $\delta=0.5$; $LnBaCo_2O_{5.5}$ (type-II) (Taken from Ref. 12 & 16).

## MAGNETIC AND ELECTRON TRANSPORT PROPERTIES

There are two characteristic structural distortions which influence the physical properties of doped rare earth cobaltites. The first one is cooperative tilting of the $CoO_6$ octahedra which is essentially established due to doping effect. This distortion is a consequence of the mismatch of the ionic radius and various factors as discussed in the previous section. The other distortion arises from the *Jahn-Teller (JT)* effect due to $Co^{3+}$ ion, which distorts the $CoO_6$ octahedra in such a way that there are long and short *Co-O* bonds (Figure 7). This occurs below a characteristic temperature for particular compounds, as for instance at *180K* for the disordered $La_{0.5}Ba_{0.5}CoO_3$ cobaltite [8]. This is well understood in terms of crystal field theory which describes how the *d*-electron of transition metal ions is perturbed by the chemical environment. The most effective distortion is the basal plane distortion (called $Q_2$ mode), with one diagonally opposite oxygen-pair displaced outwards and the other pair displaced inward. It is well established that a *JT* distortion involving a displacement of oxygen ions $\geq 0.1 Å$ can split the $e_g$-band of the cobaltites (which forms the conduction band) and opens a gap at the *Fermi* level. The magnitude of the crystal field splitting of *d*-orbital determines whether the *Co*-ion occurs in the low-spin (*LS*), intermediate-spin (*IS*) or high-spin (*HS*) configuration. Figure 7, shows a scheme of the band diagram of $LaCoO_3$ to elucidate how the *JT* distortion splits the conduction band and makes the material insulating. The octahedral ligand environment around *Co*-ion splits the five *d*-orbitals into $t_{2g}$-triplet ($d_{xy}$, $d_{yz}$ and $d_{zx}$) and $e_g$-doublet ($d_{x^2-y^2}$ and $d_{z^2}$) state. In this system, the resulting crystal-field splitting, $\Delta_{cf}$, between $t_{2g}$ and $e_g$ orbital is around *2.06 eV* as reported by *Korotin et al.* [22] for theoretical observation, although experimentally obtained values are around *1.2 eV* and *0.9eV* [22]. Further splitting of the $e_g$ orbitals due to the *JT* effect opens a gap at the *Fermi* level. The intra-atomic exchange energy responsible for *Hund's* highest multiplicity rule, $\Delta_{ex}$ (or $J_H$), is smaller than $\Delta_{cf}$ i.e. $\Delta_{ex} < \Delta_{cf}$ for $Co^{3+}$ ion. Therefore *Co*-ions are always in low spin state below *100 K* for $LaCoO_3$ [22]. This perturbation induced electronic spin-state transition in rare earth cobaltites has been of great interest in recent years. The thermally driven spin-state transition in cobaltites is a consequence of the subtle interplay between the crystal field splitting ($\Delta_{cf}$) and the *Hund's* coupling energy ($\Delta_{ex}$). The $\Delta_{cf}$ usually decreases as the temperature is increased, whereas $\Delta_{ex}$ is insensitive to temperature since it is an atomic quantity. The spin-state of undoped $LaCoO_3$ ($Co^{3+}$ ion) exhibits a gradual crossover with increasing temperature from the *LS* state ($t_{2g}^6 e_g^0$; $S = 0$) to *IS* state ($t_{2g}^5 e_g^1$; $S = 1$) at around *100 K* and finally to *HS* state ($t_{2g}^4 e_g^2$; $S = 2$) [22]. This results from the competition of the crystal field with energy $\Delta_{cf}$ ($t_{2g}$-$e_g$ splitting) and the interatomic (*Hund*) exchange energy $\Delta_{ex}$, leading to redistribution of electrons between $t_{2g}$ and $e_g$ levels.

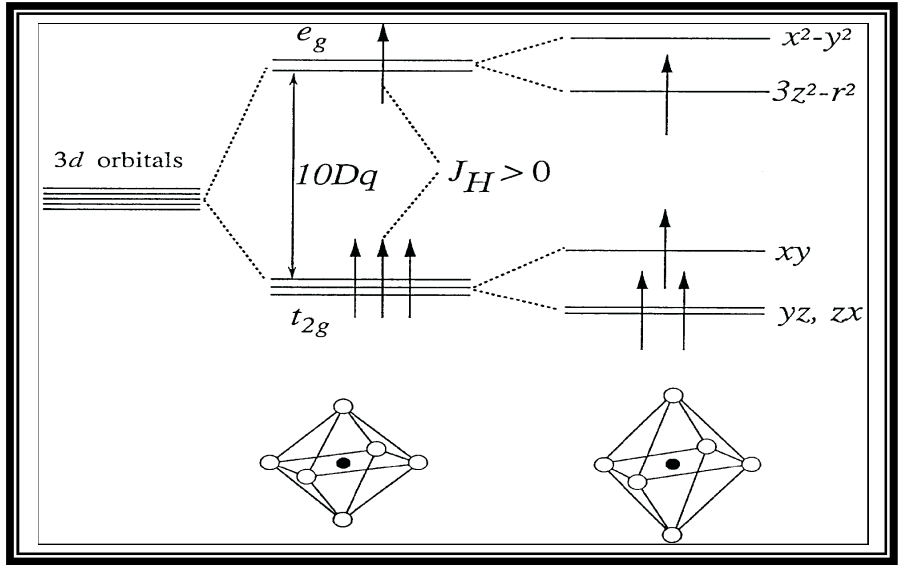

Figure 7. Field splitting of the five-fold degenerate atomic 3*d* levels for perovskite cobaltites.

In $Ln_{1-x}Ak_xCoO_3$ the doping by $Ak^{2+}$ cation induces the formal presence of $Co^{4+}$, which slightly reduces the *JT* distortion leading towards cubic structure. The *JT* distortion of $CoO_6$ octahedra has been reported by *Fauth et al.* [8], which is favored due to the *IS* state of $Co^{3+}$ ($t_{2g}^5 e_g^1$) and $Co^{4+}$ ($t_{2g}^4 e_g^1$). The transfer interaction of $e_g$-electrons is greater in rhombohedral or pseudocubic phase than in the orthorhombic phase because the ⟨Co-O-Co⟩ bond angle becomes closer to *180°*. Clearly, the doping effect or the presence of $Co^{4+}$ ion plays an important role in this material to provide the FM and metallic behavior by suppressing the *JT* distortion. The electron transport properties of cobaltites also depend on oxygen stoichiometry, and belong to the class either of metals or of semiconductors. Doped cobaltites are mixed valence materials, which mean that the Co-ions can carry different charges in the same site. The ratio between different Co-ion configurations is also determined by the oxygen stoichiometry as well as by the doping concentration, following the charge neutrality condition. Changes in the crystallographic and magnetic structures as well as the other physical properties can be induced by varying several parameters of either intrinsic hole/electron (doping, oxygen content) or extrinsic nature (temperature, pressure, magnetic field). Since only small changes of parameters (intrinsic/extrinsic) can cause a structural, magnetic or electronic transition, hence a variety of contradictory models or results exist for cobaltites. Detailed discussions are therefore required to clarify the influence of each parameter on the physical properties of the perovskite cobaltites. In this article, special emphasis has been given on the cationic ordering and oxygen stoichiometry onto the crystallographic, magnetic and electronic properties of few particular cobaltite systems. The magnetic structures, especially of the layered cobaltite systems with oxygen content of $O_{5.5}$ per unit formula, are still debated for some of the reported compounds. The determination of the magnetic structure is a complex task contrary to other materials, because the *Co*-ion in these materials can be in different spin states as discussed earlier.

## I. Disordered Cobaltites

Disordered rare earth cobaltites $Ln_{1-x}Ak_xCoO_3$ have been investigated for several years due to their novel magnetic and electronic properties which include temperature-induced spin-state transitions, cluster-glass like behavior, electronic phase separation, magnetoresistance (*MR*) and so on [6-8, 11, 13]. The physical properties of perovskite cobaltites are sensitively dependent on the doping concentration of the rare-earth site. Doping brings up mixed valences in the *Co*-ions due to charge neutrality such as $(Ln^{3+}Ak^{2+})(Co^{3+}Co^{4+})O_3$. Therefore, substitution of $Ln^{3+}$ by $Ak^{2+}$ in $Ln_{1-x}Ak_xCoO_3$ will favor the transformation of $Co^{3+}$ into $Co^{4+}$ in same ratio of doping, as a result $Co^{3+}$ and $Co^{4+}$ will interact ferromagnetically obeying the *Zener* double-exchange (*DE*) mechanism [23]. The simultaneous observation of ferromagnetism and metallicity in cobaltites is explained by this mechanism, where the hopping of an electron from $Co^{3+}$ to $Co^{4+}$ via oxygen ion, i.e. where the $Co^{3+}$ and $Co^{4+}$ ions exchange takes place. The integral defining the exchange energy in such a system is non-vanishing only if the spins of the two *d*-orbitals are parallel. That is the lowest energy of the system is one with a parallel alignment of the spins on the $Co^{3+}$ and $Co^{4+}$ ions. Due to this, the spins of the incomplete *d*-orbitals of the adjacent *Co*-ion are accompanied by an increase in the rate of hopping of electrons and therefore by an increase in electrical conductivity. Thus, the mechanism which leads to enhanced electrical conductivity requires a *FM* coupling. On the other hand, $Co^{3+}$-$Co^{3+}$ and $Co^{4+}$-$Co^{4+}$ couple antiferromagnetically due to super-exchange interactions. Super-exchange interaction generally occurs between localized moments of ions in insulators. *Goodenough et al.* pointed out that the *FM* interaction is governed not only by the *DE* interaction, but also by the nature of the super-exchange interactions [23]. Whether the ferromagnetism in cobaltites (similar to manganites) is mediated by a *DE* mechanism or not is clearly not understood at present. However, the absence of half filled $t_{2g}$ orbitals is providing core spin and strong *Hund's* rule coupling, unlike manganites, making this mechanism less feasible. It seems, the *FM*-metallic phase in cobaltites can be explained by the *Zener-DE* mechanism whereas super-exchange will fit for insulating state. Hence, there will be always a competition between these two interactions to dominate one over another giving rise to a tendency of electronic phase separation in the system [1]. The growth of interest in perovskite cobaltites is due to the expectation that, in addition to the lattice, charge and spin degrees of freedom found in many other transition metal oxides, the cobalt oxides also display a degree of freedom in the "spin-state" at the *Co*-site.

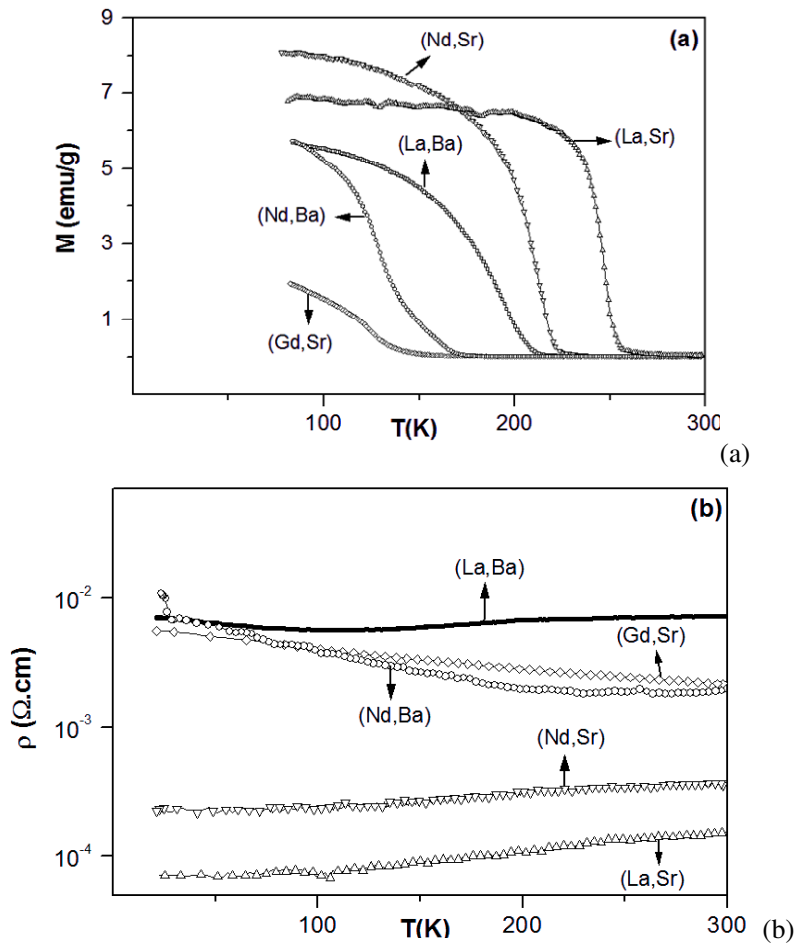

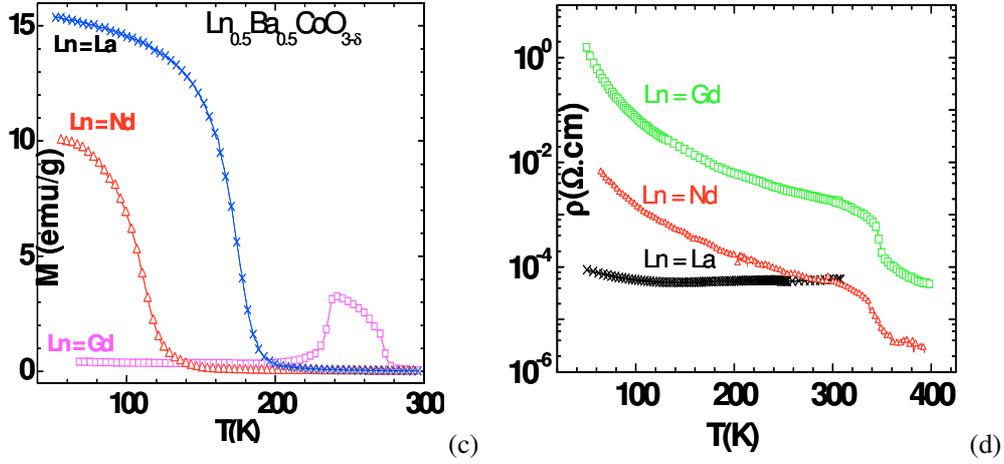

Figure 8. Temperature dependent Magnetization and Resistivity for disordered cobaltites $Ln_{0.5}Ak_{0.5}CoO_3$ (Taken from Ref. 13 & 18).

The physical properties of the cobaltites are sensitive to doping concentration at the rare earth site. Accordingly, the other parameters such as the average radius of the *A*-site cation, $\langle r_A \rangle$, and size-disorder parameter, $\sigma^2$, also vary due to doping at the perovskite *A*-site. These parameters crucially control the physical properties of disordered cobaltites. Disordered cobaltites of the type $Ln_{0.5}Ak_{0.5}CoO_3$, especially those with *Ln*= La, Nd, Gd and *Ak* = Ba, Sr are *FM*, many of them showing a metallic behavior as shown in Figure 8 [13, 18, 19]. These properties arise because of the major influence of $Co^{3+}$-O- $Co^{4+}$ interactions in these cobaltites. The *FM* $T_C$ increases with the increase in the size of the *A*-site cations, $\langle r_A \rangle$. In the case of *Ak* = Ba, ferromagnetism occurs when *Ln* = La ($T_C$ ~ *190 K*) and Nd ($T_C$ ~ *130 K*), but for *Ln* = Gd, the material shows unusual magnetic behavior (Figure 8c). Furthermore, $Gd_{0.5}Ba_{0.5}CoO_3$ is an insulator and exhibits an electronic transition around *350 K* with cationic ordering [13], whereas $La_{0.5}Ba_{0.5}CoO_3$ is metallic below *FM* $T_C$ (Figure 8d). $Gd_{0.5}Ba_{0.5}CoO_3$ which is charge-ordered at room temperature, shows a *FM*-like feature around *280 K*, without reaching a saturation value of the magnetic moment or the highest moment achieved is rather low even in the higher applied field conditions [13, 18]. The magnetic transition around *280 K* in $Gd_{0.5}Ba_{0.5}CoO_3$, has been classified in the literature as *FM* or meta-magnetic, and other cobaltites of this family such as $La_{0.5}(Nd_{0.5})Ba_{0.5}CoO_3$ have not been distinguished, and have all been treated as *FM* transitions. There are, however, considerable differences amongst these cobaltites. The magnetic transitions in $La_{0.5}(Nd_{0.5})Ba_{0.5}CoO_3$ are distinctly *FM*, showing a sharp increase in magnetization at $T_C$, and the value is rather low (*130-190 K*) [13, 18]. $Gd_{0.5}Ba_{0.5}CoO_3$, with a much smaller *A*-site cation, should have been associated with an even lower $T_C$. In Figure 9, we present the temperature variation of the magnetization of a few compositions of $Ln_{0.5}Ak_{0.5}CoO_3$. The $T_C$ values are plotted against $\langle r_A \rangle$ from the literature data, increasing up to $\langle r_A \rangle$ value of *1.40 Å* and decreasing thereafter. The decrease in $T_C$ for $\langle r_A \rangle$ *1.40 Å* is likely to arise from the size-disorder. Indeed, the cations size mismatch, $\sigma^2$, is known to play an important role in determining the properties of rare earth cobaltites [13]. It appears that the large value of $\sigma^2$ in the $Gd_{0.5}Ba_{0.5}CoO_3$ (*0.033 Å$^2$*) compared to that of $La_{0.5}Ba_{0.5}CoO_3$ (*0.016 Å$^2$*) could be responsible for the absence of ferromagnetism and metallicity in the former [13, 18].

To understand the role of cationic size, we will now discuss the magnetic and electrical properties of several series of cobaltites. The temperature dependent magnetization and resistivity behavior of $Gd_{0.5-x}Nd_xBa_{0.5}CoO_{3-\delta}$ series is presented Figure 10. With increasing the *Nd* substitution or the *<r$_A$>* value, the evolution of ferromagnetism is observed. It is interesting to note that the *280 K* magnetic transition of $Gd_{0.5}Ba_{0.5}CoO_{2.9}$ disappears even when $x \geq 0.1$. Moreover, for *x = 0.1*, a complex behavior is observed with a magnetic transition around *220 K*. The *x = 0.3* composition shows a weak increase in magnetization value around *125 K*, and the effect is more prominent for *x = 0.4*. Similarly, for $Gd_{0.5-x}La_xBa_{0.5}CoO_{3-\delta}$, there is no clear *FM* transition in the *200-280 K* region for *0.1 < x < 0.25*. A distinct *FM* transition occurs at *x = 0.5* in the case of *Nd*, and at *x = 0.4* in the case of *La*. It is interesting that the *FM* characteristics start emerging at low temperatures (*< 150 K*) in these cobaltite compositions around a ⟨r$_A$⟩, value of *1.30 Å*. Clearly with increasing *x*, the size of ferromagnetic clusters increases, eliminating the phase separation at lower value of *x*, caused by size disorder. It is noteworthy that in the $Ln_{0.7}Ca_{0.3}CoO_3$ (*Ln = La, Pr, Nd*) series the spin glass like behavior has been reported at low temperatures, the system approaching toward spin glass state [11] for the higher size disorder. Spin glass behavior is also reported for $La_{1-x}Sr_xCoO_3$ (*x < 0.1*), but with increase in *x* ferromagnetism manifests itself [13].

$Gd_{0.5-x}Nd_xBa_{0.5}CoO_{3-\delta}$ exhibit an insulating behavior throughout the temperature range, but the electrical resistivity decreases significantly with increase in *x*, particularly *x = 0.5* composition exhibiting the lowest resistivity (Figure 10). Similarly for *Ln = La*, the resistivity value decreases with increasing *x*, becoming metallic for *x = 0.5*. All the other compositions are semiconducting. Considering that with increase in *x*, there is significant increase in *<r$_A$>* values in these disordered $Gd_{0.5-x}Ln_xBa_{0.5}CoO_{3-\delta}$ (*Ln = La, Nd*) series of cobaltites, the changes observed can essentially be attributed to the effects of cation size which indirectly control the electronic band width and as a result the energy band gap varies.

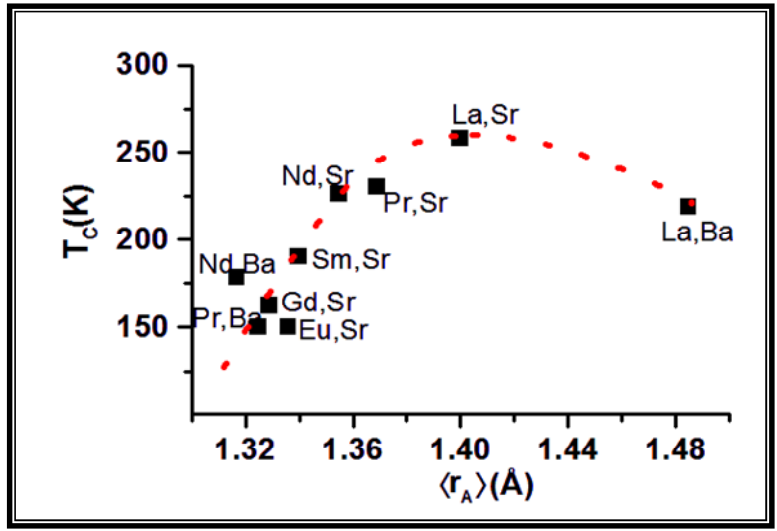

Figure 9. Variation of FM T$_C$ with <r$_A$> for disordered cobaltites $Ln_{0.5}Ak_{0.5}CoO_3$ (Taken from Ref. 13).

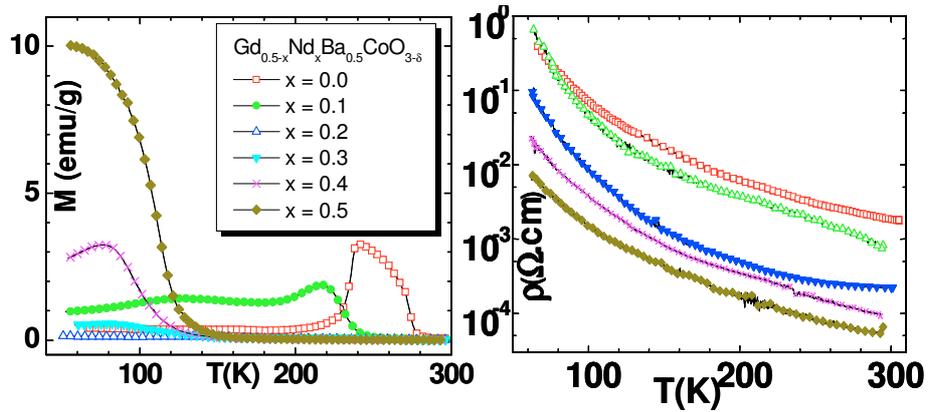

Figure 10. Temperature dependent Magnetization and Resistivity for $Gd_{0.5-x}Nd_xBa_{0.5}CoO_3$.

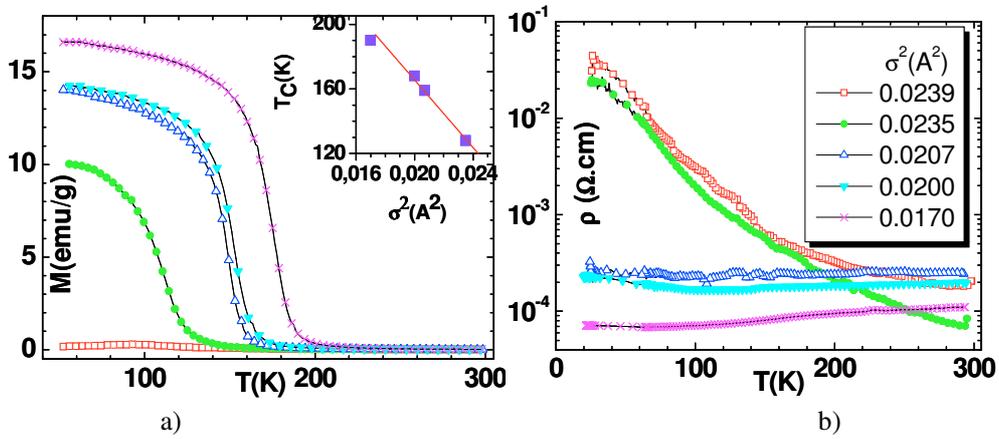

Figure 11. Temperature dependent Magnetization and Resistivity for a fixed $<r_A>$ of $1.317$ Å, inset figure (a) shows the $T_C$-$\sigma^2$ plot (Taken from Ref. 18).

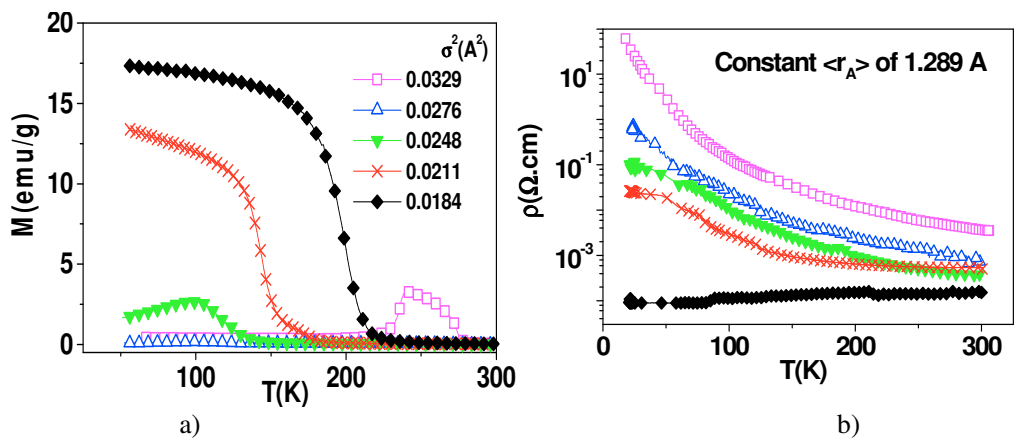

Figure 12. Temperature dependent Magnetization and Resistivity for a fixed $<r_A>$ of $1.289$ Å (Taken from Ref. 18).

In order to understand the role of size-disorder due to cation size mismatch, $\sigma^2$, we have discussed two series of cobaltites with fixed $\langle r_A \rangle$ values of *1.317* and *1.289 Å*, corresponding to those of $Nd_{0.5}Ba_{0.5}CoO_3$ and $Gd_{0.5}Ba_{0.5}CoO_3$ respectively. The data, for a fixed $\langle r_A \rangle$ of *1.317 Å*, show that the *FM* $T_C$ decreases with increasing $\sigma^2$ (inset Figure 11a), eventually destroying ferromagnetism at a high value of $\sigma^2$ ($\approx 0.024\ Å^2$). Similarly, the data in Figure 12, for a fixed $\langle r_A \rangle$ of *1.289 Å*, are interesting. It is observed that with decreasing $\sigma^2$, the magnetic behavior of this system changes markedly. Thus, when $\sigma^2 = 0.028\ Å^2$, there is no such magnetic anomaly as observed in the case of $Gd_{0.5}Ba_{0.5}CoO_{2.9}$ at *280 K*. When $\sigma^2 = 0.021\ Å^2$, we observe a *FM* transition with a $T_C \sim 160\ K$ and for $\sigma^2 = 0.018\ Å^2$, the $T_C$ value reaches *220 K*, a value higher than that of $La_{0.5}Ba_{0.5}CoO_3$. These data clearly demonstrate that the absence of prominent ferromagnetism in $Gd_{0.5}Ba_{0.5}CoO_3$, and the unusual magnetic properties, like magnetic anomaly around *280 K*, is almost entirely due to the disorder arising from the cation size mismatch. Such size-disorder can give rise to electronic phase separation as reported for other cobaltites [18]. The electrical resistivity of these series of cobaltites corroborates the results from the magnetic measurements. In Figure 12b, we present the electrical resistivity data to demonstrate how the resistivity increases with increase in $\sigma^2$. Interestingly, disorder-induced insulator-metal transitions are noticed in both the series of cobaltites and the compositions with $\sigma^2 < 0.02\ Å^2$ showing metallic behavior. While disorder-induced meal-insulator transitions are common, size variance-induced insulator-metal transitions are indeed novel.

Further support of the previous discussion that cation size-disorder crucially determines the properties of $Gd_{0.5}Ba_{0.5}CoO_{2.9}$ is provided in this section by the study of $Gd_{0.5}Ba_{0.5-x}Sr_xCoO_{3-\delta}$ series of cobaltites as reported in the literature [18]. Here, $x = 0.5$ composition, corresponding to $Gd_{0.5}Sr_{0.5}CoO_3$, has a smaller $\langle r_A \rangle$ than $Gd_{0.5}Ba_{0.5}CoO_{2.9}$, and yet it shows ferromagnetic features. The *280 K* magnetic anomaly of $Gd_{0.5}Ba_{0.5}CoO_3$ disappears even when $x = 0.1$ and the apparent $T_C$ increases with increasing *x* in this series. This behavior is clearly due to size disorder effect, since $\sigma^2$ decreases with increase in *x*. Accordingly, the present system exhibits an insulator-metal transition with increase in *x* or decrease in $\sigma^2$. It appears that a $\sigma^2$ value larger than *0.02 Å²* generally destroys ferromagnetism in the cobaltites and changes the metal into an insulator as discussed earlier.

In view of the interesting magnetic and electrical properties of the disordered $Ln_{0.5}Ba_{0.5}CoO_3$ cobaltites, we have also presented a comparative study of *Sr*-doped disordered cobaltites $Ln_{0.5}Sr_{0.5}CoO_3$ with *Ln = La, Pr, Nd,* and *Gd*. Magnetic properties of $La_{0.5}Sr_{0.5}CoO_3$ are fairly well understood. Although it shows a sharp *FM* $T_C$ around *240 K*, it exhibits a significant divergence between the *FC* and *ZFC* magnetization data (Figure 13a) and also shows a frequency-dependent *AC* susceptibility maximum around *165 K*, suggesting a glassy nature [19]. It was realized a few years ago that $La_{0.5}Sr_{0.5}CoO_3$, which was considered to be a good ferromagnetic metal, was actually a magnetic cluster-glass with some long-range magnetic ordering wherein frustration arose from inter-cluster interactions at low temperatures [6]. The magnetic behavior of these systems has generally been interpreted in terms of short-range magnetic ordering. Cluster-glass behavior occurs at a high concentration of *Sr* (*x > 0.3*), where the coalescence of short-range ferromagnetic clusters is proposed to occur [6, 13]. $^{139}La\ NMR$ studies confirm the coexistence of ferromagnetic, paramagnetic, and cluster-glass phases in $La_{1-x}Sr_xCoO_3$ [6, 13]. Electronic phase separation in the $La_{1-x}Sr_xCoO_3$ systems is associated with the formation of isolated nanoscopic ferromagnetic clusters. Thus, the cobaltites comprise ferromagnetic clusters, paramagnetic matrices and spin-glass-like phases, all contributing to the glassy magnetic behavior. More importantly, this glassy ferromagnetism is accompanied by phase separation wherein the ferromagnetic clusters exist within a antiferromagnetic (*AFM*) matrix. In these cobaltites, the *FM* phase is conducting and the *AFM* phase is insulating. Depending on *x* or the carrier concentration, we can have a situation such as that shown in Figure 14. The phase separation scenario here is somewhat complex because the transition from the metallic to the insulating state is not sharp. In the presence of *Coulomb* interaction, the microscopically charged inhomogeneous state is stabilized, giving rise to clusters of one phase embedded in another. Electronic phase separation with phases of different charge densities is generally expected to give rise to nanometer scale clusters [19]. This is because large-phase separated domains would break up into small pieces because of *Coulomb* interactions. The shapes of these pieces could be droplet or stripes (see Figure 14a-d). The domains of the two phases can also be sufficiently large to give rise to well-defined signatures in neutron scattering or diffraction. One can visualize phase separation arising from disorder as well. The disorder can arise from the size mismatch of the *A*-site cations in the disordered perovskite structure [1]. Such phase separation is reported for *(La$_{1-y}$Pr$_y$)$_{1-x}$Ca$_x$MnO$_3$* system in terms of a metal-insulator transition induced by disorder [1]. The size of the clusters depends on the magnitude of disorder. The smaller the disorder, the larger would be the size of the clusters. This could be the reason why high *MR* occurs in systems with small disorder.

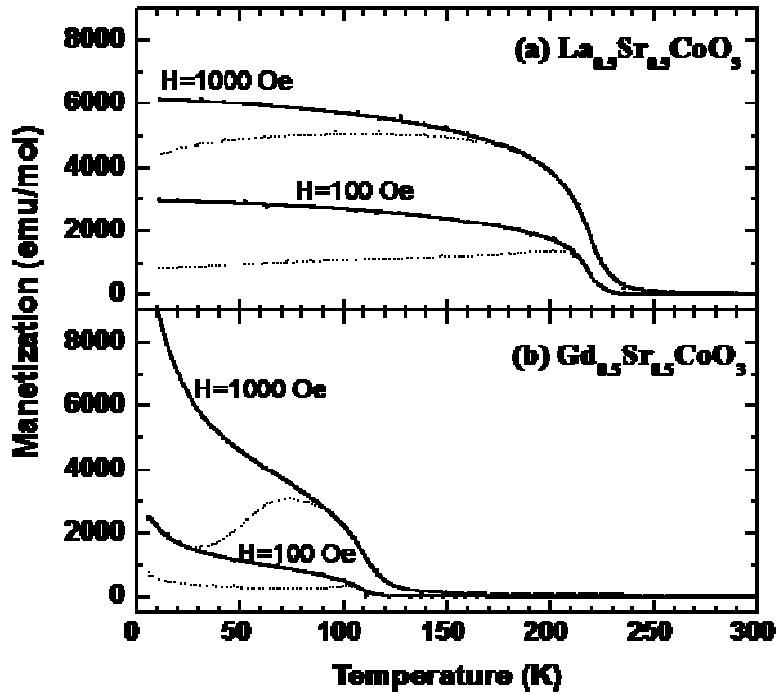

Figure 13. Temperature dependent FC-ZFC Magnetization for (a) *La$_{0.5}$Sr$_{0.5}$CoO$_3$* and (b) *Gd$_{0.5}$Sr$_{0.5}$CoO$_3$* (Taken from Ref. 19).

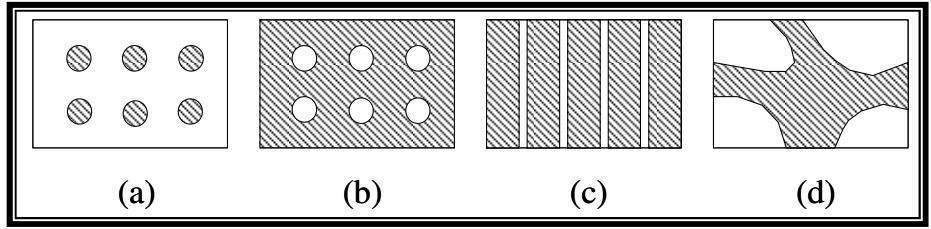

Figure 14. Schematic representation of electronic phase separation (Taken from Ref. 1).

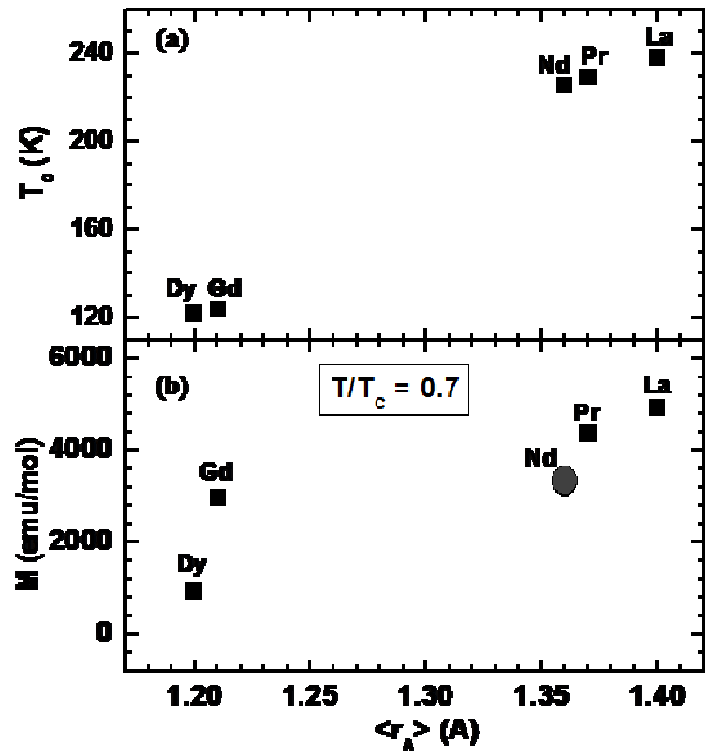

Figure 15. Variation of (a) FM $T_C$ and (b) Magnetization with $<r_A>$ for disordered cobaltites $Ln_{0.5}Sr_{0.5}CoO_3$ at *120 K* (Taken from Ref. 19).

Electronic phase separation in the $La_{1-x}Sr_xCoO_3$ systems is associated with the formation of isolated nanoscopic *FM* clusters [6]. Evidence for the occurrence of phase separation in $La_{0.5}Sr_{0.5}CoO_3$ is provided by the *NMR* studies [6]. The *Mössbauer* spectra reported by *Bhide et al.* [13] show the presence of a paramagnetic signal in addition to the six-finger pattern due to the *FM*-type species over a range of compositions and a wide range of temperatures. The temperature variation of the ferromagnetic-paramagnetic (*FM/PM*) ratio was reported recently for $La_{0.5}Sr_{0.5}CoO_3$ [19], showing an increasing ratio with decreasing temperature, as one would expect. While the *FM/PM* ratio increases with a decrease in temperature, the *PM* phase continues to exist well below $T_C$ in $La_{0.5}Sr_{0.5}CoO_3$. Clearly, this observation is a direct evidence for phase separation in $La_{0.5}Sr_{0.5}CoO_3$. It is of interest to examine the glassy phases and phase separation occurring in the other rare earth cobaltite compositions of the type $Ln_{0.5}Sr_{0.5}CoO_3$. The temperature-dependent *DC* magnetization data of $Gd_{0.5}Sr_{0.5}CoO_3$ exhibits divergence in the *ZFC* and *FC* behavior just as in magnetically disordered systems. What is particularly noteworthy is that the magnetization value in $Gd_{0.5}Sr_{0.5}CoO_3$ is much smaller than the *Pr* and *La* counterparts below their $T_C$'s, especially at temperatures below the relatively sharp transition around *110 K*. Since the carrier concentration or the $Co^{3+}/Co^{4+}$ ratio remains constant in these cobaltites, the only possible explanation for such a decrease in magnetization is that the proportion of *PM* species relative to that of the *FM* species increases with the decrease in the size of the rare earth ion. It is entirely understandable that the proportion of the clusters responsible for ferromagnetism decreases with decrease in $<r_A>$. In the $Ln_{0.5}Sr_{0.5}CoO_3$ series, the *FM* $T_C$ decreases with decreasing $<r_A>$ (Figure 15) which is a well-established behavior for disordered cobaltites [13, 19]. Since the magnetic moment below $T_C$ is related to the proportion of the *FM* clusters, the variation of magnetization with $<r_A>$ at *120 K* in a series of $Ln_{0.5}Sr_{0.5}CoO_3$ compounds is shown in Figure 15. We have presented the size disorder effect in the $Ln_{0.5}Sr_{0.5}CoO_3$ system as well. For this purpose, the data were presented for fixed $<r_A> = 1.196$ Å as equal to that of $Dy_{0.5}Sr_{0.5}CoO_3$ and varying the $\sigma^2$ values for different cobaltite compositions. The plot of $T_C$ values against $\sigma^2$ is shown in Figure 16. The *FM* $T_C$ increases with a decreasing $\sigma^2$ and the $T_C^0$ value corresponding to the disorder-free case ($\sigma^2 =0.0$) is *217 ± 2 K*. The linear relation between $T_C$ and $\sigma^2$ gives a slope of *9667 ± 192 K Å$^{-2}$*, which is comparable to that reported in the literature [13, 18]. Thus, it can be concluded for disordered cobaltites that, increasing size disorder favors electronic phase separation, giving rise to magnetic clusters of different size ranges, while decreasing size disorder increases the *FM/PM* ratio.

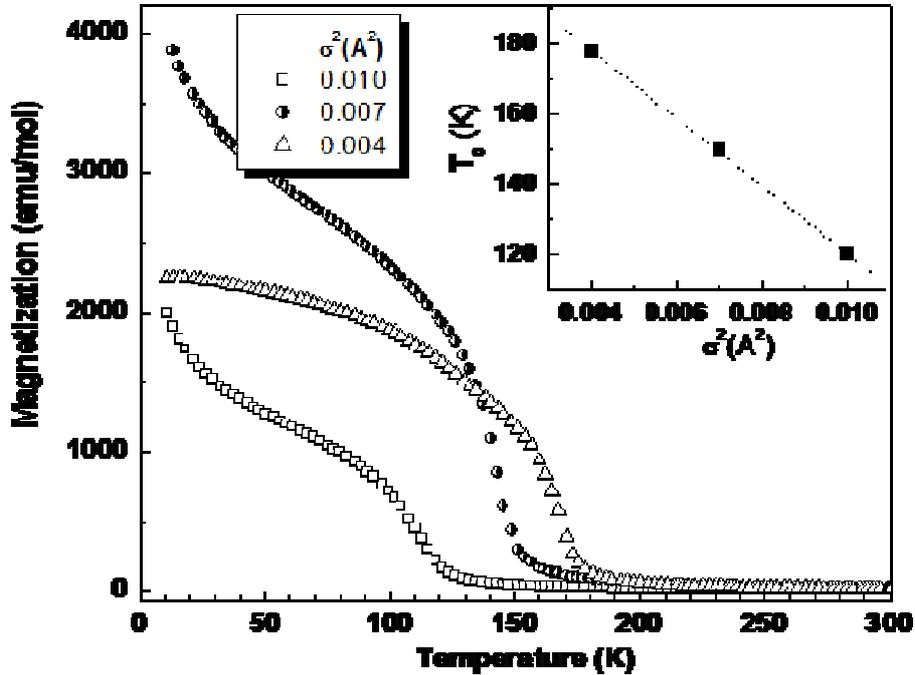

Figure 16. Magnetic properties for a fixed $<r_A>$ of *1.196 Å* (same as $Dy_{0.5}Sr_{0.5}CoO_3$) (Taken from Ref. 19).

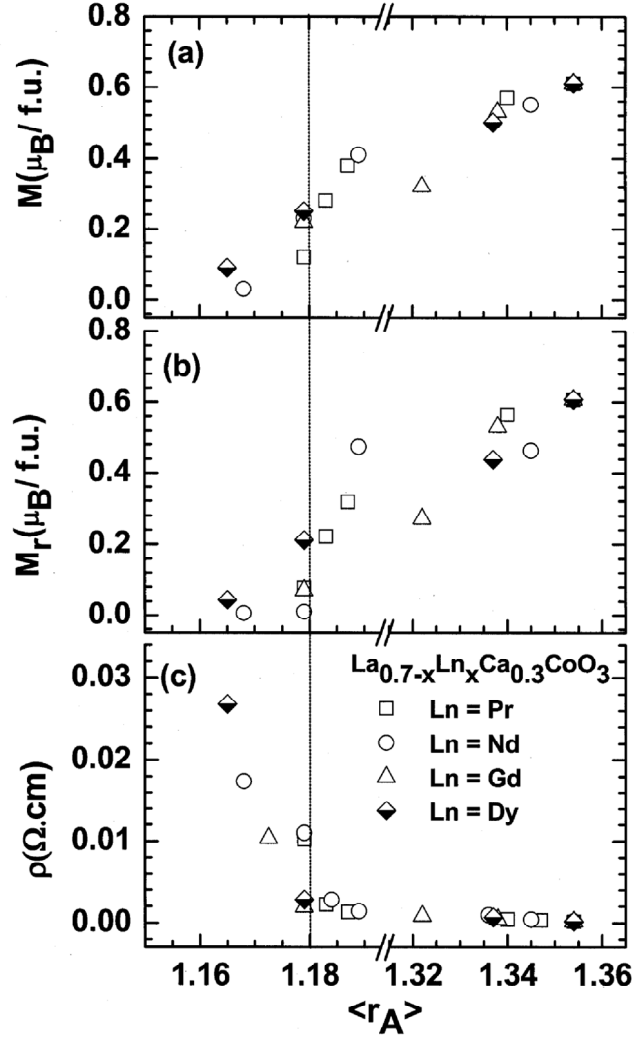

Figure 17. Magnetization and Resistivity behavior for disordered $La_{0.7-x}(Pr/Nd/Gd/Dy)_xCa_{0.5}CoO_3$ series with the variation of $<r_A>$ at *50 K* (Taken from Ref. 11).

The value of *MR* for disordered cobaltites (even at $T_C/T_{IM}$) is much smaller in magnitude than for the manganites [7]. However, in disordered cobaltites the large and negative *MR* has been reported in the insulating compositions of $La_{0.5}Sr_{0.5}CoO_3$ [7]. In such system, the maximum *MR* is observed where it shows *SG*-like behavior [7, 13]. Disordered cobaltites of the type $Ln_{1-x}Ca_xCoO_3$ show no long-range ferromagnetism or insulator-metal transition, instead they exhibit electronic phase separation and/or glassy magnetic behavior at low temperatures [11, 13]. Studies on $La_{1-x}Ca_xCoO_3$ samples have suggested that there are no major differences from the *Sr*-doped system. Ferromagnetism is observed in both systems, with the Curie temperature being lower in *Ca*-doped materials at a fixed doping level. Magnetic and electron transport properties of different rare earth cobaltites have been investigated to examine the effect of $<r_A>$, and $\sigma^2$ on these systems as presented in Figure 17.

Thus, while $La_{0.7}Ca_{0.3}CoO_3$ ($\langle r_A \rangle$, = 1.354 Å) shows glassy ferromagnetism associated with metallicity at low temperature, $Ln_{0.7}Ca_{0.3}CoO_3$ with a smaller $\langle r_A \rangle$, of 1.179 Å (Ln = Pr) and 1.168 Å (Ln = Nd) shows no long-range ferromagnetism or insulator-metal transition [11]. Instead, the latter two systems exhibit electronic phase separation and/or spin-glass like behavior at low temperatures. The electronic phase separation and associated magnetic properties of $Pr_{0.7}Ca_{0.3}CoO_3$ and $Nd_{0.7}Ca_{0.3}CoO_{2.95}$ arise because of the small average size of the A-site cations [11]. In these two cobaltites, the average radius (for orthorhombic structure) is less than *1.18 Å*, which is the critical value only above which long-range ferromagnetism manifests itself. A detailed study on disordered rare earth cobaltites has shown the occurrence of electronic phase separation and glassy magnetic behavior for small $\langle r_A \rangle$, and a large $\sigma^2$ value [11].

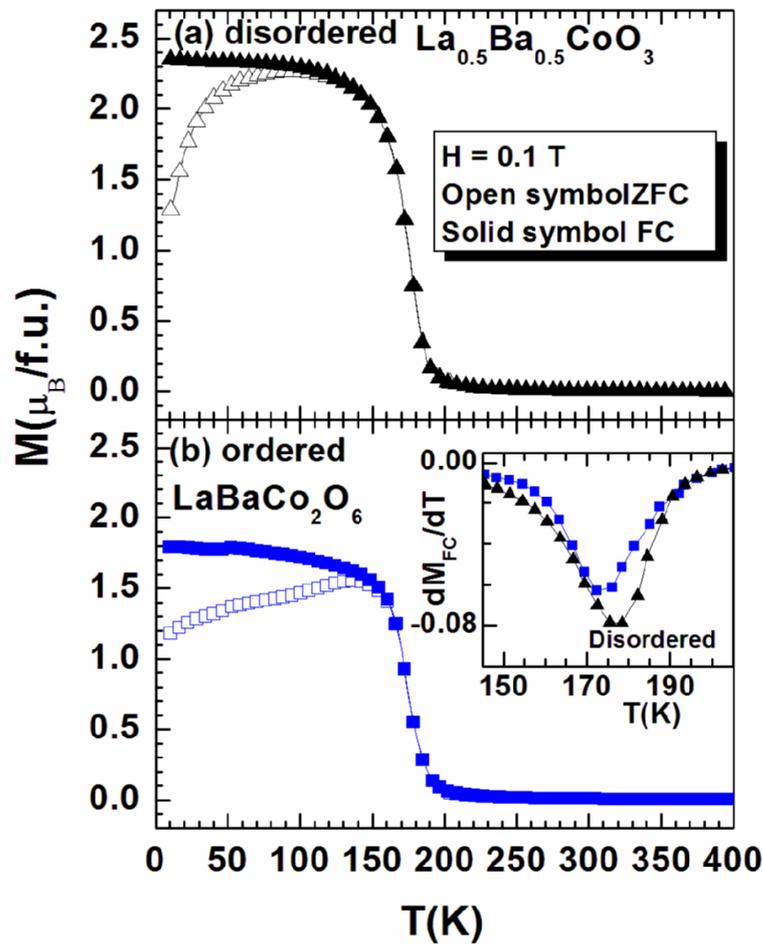

Figure 18. Temperature dependence ZFC (open symbol) and FC (solid symbol) Magnetization (H = 0.1 Tesla) for (a) disordered $La_{0.5}Ba_{0.5}CoO_3$ and (b) ordered $LaBaCo_2O_6$. The inset figure shows $dM_{FC}/dT$ vs temperature plot.

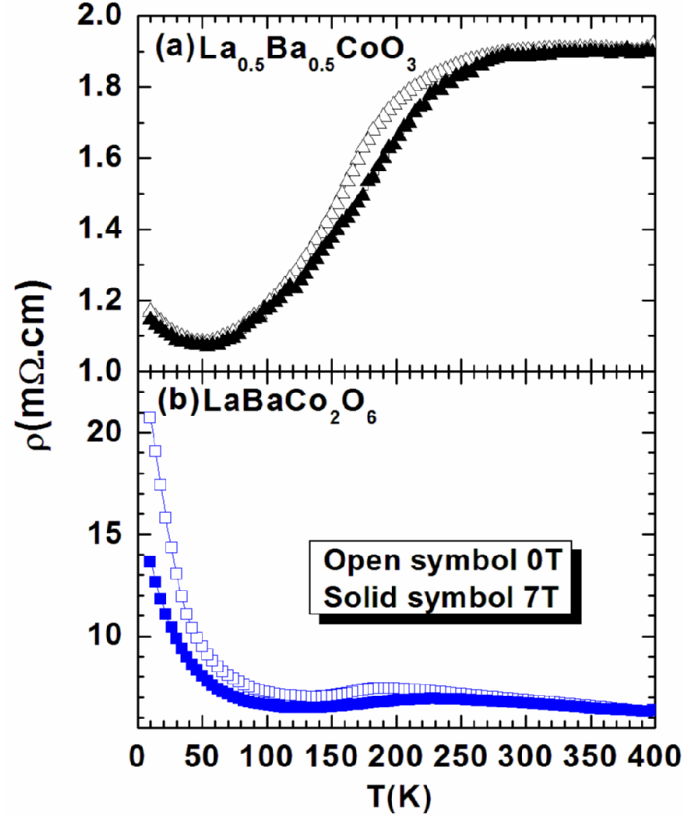

Figure 19. Temperature dependence electrical resistivity, ρ, for (a) disordered $La_{0.5}Ba_{0.5}CoO_3$ and (b) ordered $LaBaCo_2O_6$ cobaltites (Taken from Ref. 12).

## II. Ordered Cobaltites

The cationic order-disorder phenomena in the perovskite cobaltites do not affect much their *PM-FM* transition temperature $T_C$ [9], in contrast to the ordered $LaBaMn_2O_6$ and disordered $La_{0.5}Ba_{0.5}MnO_3$ manganites [20]. However, the influence of the cationic ordering upon the $T_C$ seems to be reverse since, according to *Nakajima et al.* [9], the disordered cobaltites exhibit a higher $T_C$ of *190 K* than the ordered phase *(175 K)*. Figure 18, shows the temperature dependent magnetization measurements for the disordered $La_{0.5}Ba_{0.5}CoO_3$ and ordered $LaBaCo_2O_6$ also indeed exhibit $T_C$ of *177 K* and *174 K* respectively [12]. Moreover, other magnetic behaviors such as the field and the temperature dependent magnetization are similar to expected for *PM/FM* phases [9, 12]. The coercive field, $H_C$, for disordered $La_{0.5}Ba_{0.5}CoO_3$ and ordered $LaBaCo_2O_6$ are *0.08* and *0.05 Tesla (T)* respectively (at *10 K*); the low value of $H_C$ signifies the nature of soft *FM* material [12]. There is one more cobaltite compound reported in the literature as ordered-disordered phases, with a FM $T_C$ of 250 K in the ordered $NdBaCo_2O_6$ phase [14] whereas for disordered $Nd_{0.5}Ba_{0.5}CoO_3$ phase [13] the value is 130 K. Since those compounds are reported by two different groups so we will not elaborate our discussion on those phases for the present article.

Figure 19 shows the temperature dependence of electrical resistivity for both ordered-disordered phases in the presence and absence of an applied magnetic field of ±7 *Tesla(T)*. Note that the resistivity, $\rho(T)$, behavior of the disordered $La_{0.5}Ba_{0.5}CoO_3$ and ordered $LaBaCo_2O_6$ are different from those reported by *Nakajima et al.* [9], and corroborate the result reported by *Fauth et al.* [8] for the disordered $La_{0.5}Ba_{0.5}CoO_3$. The $\rho(T)$ curve in *10-400 K* temperature range depicts that at high temperature (*T>300K*) the disordered $La_{0.5}Ba_{0.5}CoO_3$ (Figure 19a) phase is semi-metallic, whereas the ordered $LaBaCo_2O_6$ is semiconducting down to *190 K* (Figure 19b). This feature is explained by the fact that the disordered cobaltite exhibits *180°* ⟨Co-O-Co⟩ bond angle in this temperature range, in agreement with its cubic or pseudo cubic structural symmetry, favoring a perfect overlapping of the *Co 3d* orbitals and oxygen *2p* orbitals. This is in contrast to the ordered $LaBaCo_2O_6$, where the ⟨Co-O-Co⟩ bond angles of *174°* in the equatorial planes of the $[CoO_2]_\infty$ layers are observed at room temperature [9]. In this case the conduction of charge carriers will be more favorable for linear bond angle, as a result metallic type of conductivity is noticed for the disordered cobaltite. With decreasing temperature a transition to a nearly metallic state is observed for both phases. It is characterized by a change in slope of $\rho(T)$ at $T_C$ for the disordered $La_{0.5}Ba_{0.5}CoO_3$ (Figure 19a), or by a flat maximum at $T_C$ for the ordered $LaBaCo_2O_6$ (Figure 19b). Thus, these results show that irrespectively of their structural nature, the different forms of cobaltites exhibit a *FM* metallic behavior below $T_C$. Moreover, both phases depict an upturn in the resistivity behavior at low temperature. This feature is interpreted as a weak localization contribution associated with electron-electron interaction. In the present case, the magnetoresistance measurements, that will be discussed later, suggest that the upturn is rather due to grain boundary effects.

It is now important to discuss about the oxygen deficient ordered cobaltites $LaBaCo_2O_{5.5}$, which have been of great interest due to their rich physical properties and interesting structural phenomena associated with them. In these phases the oxygen stoichiometry is "5.5", hence the average valency of cobalt ion is $Co^{3+}$ unlike the presence of mixed valences in the other two phases. This is particularly interesting because of the ordering of $Co^{3+}$ ions in two different crystallographic sites corresponding to pyramidal and octahedral oxygen coordination as discussed earlier. The magnetization and susceptibility curves versus temperature, *M (T)*, measured in the range *10-400 K*, under external fields between *0.01* and *5 T* (Figure 20) were reported recently by *Rautama et al.* [16] to be similar to those previously observed for other lanthanides [3, 14]. In the whole temperature range the system exhibits several magnetic transitions from *PM* to *FM*-like to antiferromagnetic as the temperature decreases from *400 K* to *10 K*. The sudden increase of the magnetic susceptibility at $T_C$ = *326 K* indicates a *PM/FM* transition and a sharp decrease in the magnetization at $T_N$ = *295 K* indicates a *FM/AFM* transition. The interactions between $Co^{3+}$ ions, both in pyramidal and octahedral coordination, are found to be *AFM* at low temperature. Moreover, the thermomagnetic irreversibility between *ZFC* and *FC* in the low temperature *AFM* state remains well discernible even at higher field ($H \geq 5\ T$). It is reported that a high field basically affects the *FM-AFM* competition, which in turn suppresses the magnetization drop below the $T_N$. Thus, with increasing the external magnetic field, the *FM* state (*260 K $\leq T \leq$ 326 K*) becomes more stable (indeed $T_C$ increases and the *FM* region expands in the temperature scale), but $T_N$ shifts to lower temperature. In the *AFM* phase, the nonzero value of magnetization down to low temperature signifies the presence of some kind of *FM*-like interactions, where some weak magnetic transition is evidenced. Inset of Figure 20 depicts the inverse magnetic susceptibility versus temperature plot in the temperature range of *220-400 K*. This follows a simple *Curie-Weiss* behavior in the *335 K ≤ T ≤ 400 K* range giving a *PM Weiss* temperature ($\theta_p$) of ~ *-290 K* and an effective *PM* magnetic moment ($\mu_{eff}$) of *5.27 $\mu_B$/f.u.* The large negative $\theta_p$ value for ordered *LaBaCo$_2$O$_{5.5}$* sample indicates the existence of *AFM* type interactions in the high temperature region. The obtained $\mu_{eff}$ value (*5.27 $\mu_B$/f.u.*) from the high temperature region magnetic data corresponds to a situation where the *Co$^{3+}$* ions are most probably in the *IS* state, which corroborate the results of *NPD* [16]. The sharp drop in inverse susceptibility near $T_C$, is similar to that observed for *YBaCo$_2$O$_{5.5}$* and *GdBaCo$_2$O$_{5.5}$* [21, 24].

The *FM*-like features below $T_C$ ~ *326 K*, have been confirmed by the magnetic field dependent isotherm magnetization, *M(H)*, studies at six different temperatures as shown in Figure 21. The *M(H)* curve at *300 K* shows a prominent hysteresis loop with a remanent magnetization, $M_r$, and a coercive field, $H_C$, values of *0.02 $\mu_B$/f.u* and *0.1T* respectively, indicating a *FM*-like state below $T_C$. Nevertheless, the maximum value of the magnetic moment measured in *5T* (*0.23$\mu_B$/f.u.*) at *275 K* is much smaller than the theoretical spin-only value (*4$\mu_B$/f.u.*) of *Co$^{3+}$* in *IS* state. Therefore, the *FM*-like behavior of this compound is most probably due to the canting of the magnetic spin alignment in the *AFM* phase, often observed in other systems [14]. Although there are some controversies in the literature to explain these *FM*-like features, yet this behavior is prominent for this ordered cobaltite and subject to further investigations. It is worth pointing out that at low temperature in the *AFM* state, some *FM*-like phase is still present as evidenced from the *M(H)* behavior, with a finite value of the coercive field. Interestingly, the highest moment of *0.23 $\mu_B$/f.u.* (at *275 K*) and highest coercive field value of *0.4 T* (at *245 K*) are obtained in the *AFM* region compared to the values of *0.16 $\mu_B$/f.u.* and *0.1 T* in the *FM* region (at *300 K*). For comparison the $H_C$ values have been re-plotted for lower fields at *245* and *300 K* in the inset of Figure 21. The $H_C$ values in the *AFM* region (*275-200 K*) are larger than in the *FM* region and finally at *10 K*, the *M(H)* behavior becomes linear akin to *AFM* state.

Figure 22 shows $\rho(T)$ for *LaBaCo$_2$O$_{5.5}$* in the temperature range of *10-400 K* (re-plotted *M(T)* for comparison), which is plotted for heating and cooling cycle of the measurements in the presence and absence of the external magnetic field of 7 T. The zero field $\rho(T)$ curve shows a significant change in slope corresponding to the semiconductor-semiconductor transition ($T_{SC}$) around *326 K* (Figure 22a). This type of transition is previously reported for the other series of *LnBaCo$_2$O$_{5.5}$* [14, 24, 26], and referred to as $T_{IM}$ albeit the true nature of this transition is semiconducting to semiconducting type. For *LaBaCo$_2$O$_{5.5}$* system, in contrast to a metallic behavior, the slope of the resistivity curve ($d\rho/dT$) is negative above the transition temperature ($T > T_{SC}$). Furthermore, for *LaBaCo$_2$O$_{5.5}$* there is absence of any significant change in the resistivity behavior through out the temperature range, apart from a slight decrease in the magnitude below $T_{SC}$. It is noticed that, the electronic and magnetic transition temperatures for *LaBaCo$_2$O$_{5.5}$* are almost the same ($T_C$ & $T_{SC} \approx 326$ K), in contrast to other ordered cobaltites *LnBaCo$_2$O$_{5.5}$*, which exhibit a large shift between them [14, 24, 26, 27]. Hence, the sample is magnetic-semiconductor below $T_{SC}$ and the resistivity increases exponentially with decreasing temperature.

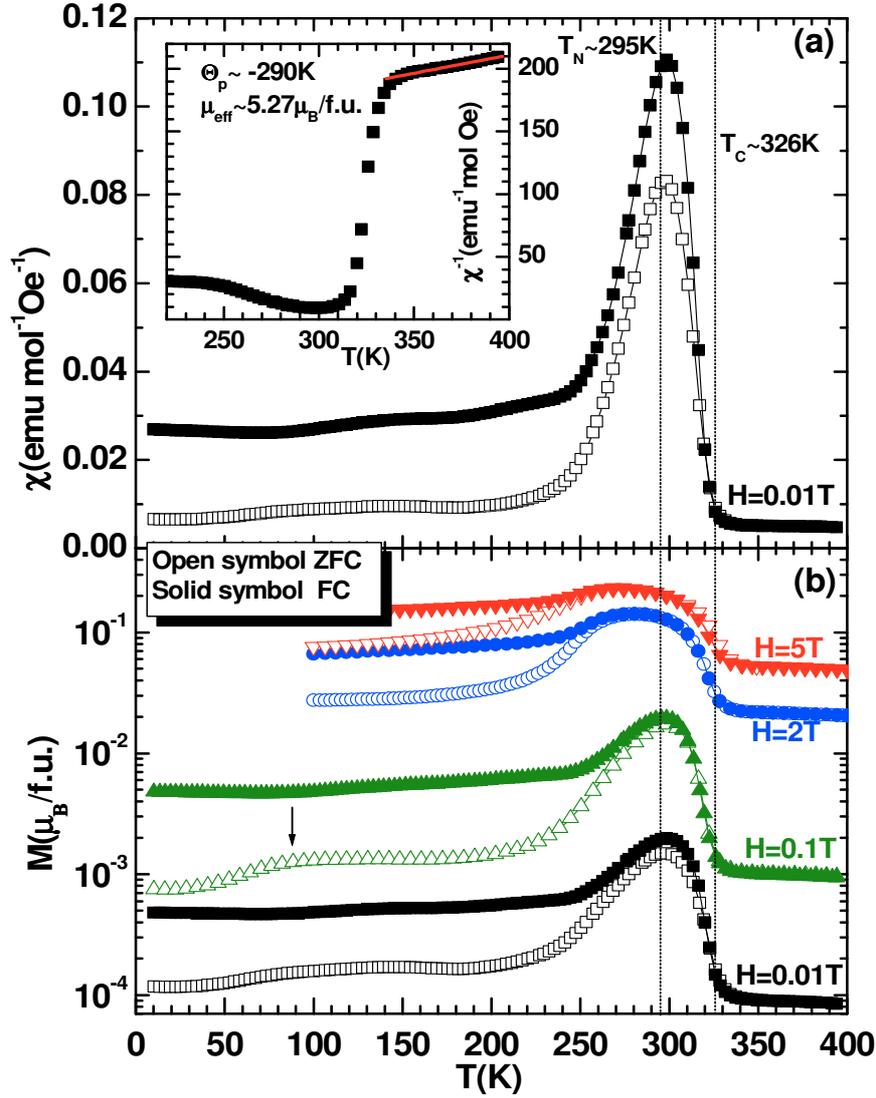

Figure 20. Temperature dependent ZFC (open symbol) and FC (solid symbol) Magnetization behavior of ordered $LaBaCo_2O_{5.5}$; (a) magnetic susceptibility, χ, under H=0.01 Tesla (inset figure shows the inverse magnetic susceptibility, $χ^{-1}$, versus temperature plot and solid line is Cuire-Weiss fitting) and (b) Magnetic moment in different magnetic fields (H=0.01, 0.1, 2 and 5 Tesla).

In general, the semiconductor or insulator like transport properties in perovskite cobaltites were characterized by three possible models [11, 25] namely thermal activation (*TA*): *log ρ α 1/T*, *Efros-Shklovskii* type hopping (*ESH*): *log ρ α $T^{-1/2}$* and *Mott's* variable range hopping (*VRH*): *log ρ α $T^{-1/4}$*. To understand the transport mechanism for $LaBaCo_2O_{5.5}$, the data have been analyzed based on these models. The zero field resistivity data is plotted in Figure 23a, which shows that the *VRH* model fits better than the other two models in *40 K ≤ T ≤ 270 K* range, and is consistent with the preceding studies on perovskite cobaltites [11, 25]. This type of transport mechanism is typical for disordered systems as shown in Figure 23b, where the charge carrier's move by hopping between localized states [11]. *Taskin et al* [26] have described the formation of localized states in terms of oxygen defects in *112*-type ordered *$GdBaCo_2O_{5.5}$*, which inevitably generate electrons or holes in the *$CoO_2$* planes. Furthermore, we consider *ESH* and *TA* models, which are expected to describe the dominant conduction process. The transport mechanism is very complicated in the whole temperature region and does not satisfy entirely any of the above models for *$LaBaCo_2O_{5.5}$*. From the nearly linear region of the *TA* model (≤ *250 K*) we have calculated the approximate activation energy ($E_a$) of around *21 meV*, yet less than the value reported for *$GdBaCo_2O_{5.5}$* [26].

The magnetoresistance (*MR*) behavior of the stoichiometric ordered-disordered phases (Figure 24) shows a clear magnetic field dependent change in the resistivity below *$T_C$*. The *MR* value is calculated as *MR (%) = [{ρ(7)-ρ(0)}/ρ(0)]x100*, where *ρ(0)* is the sample resistivity at *0 T* and *ρ(7)* under an applied field of *±7 T*. For the disordered *$La_{0.5}Ba_{0.5}CoO_3$* (Figure 24a), the maximum *MR* value is obtained around *$T_C$* and the corresponding value is indeed *7 %* at *179 K*. The ordered *$LaBaCo_2O_6$* exhibits (Figure 24b) a rather close value of *6 %* around *$T_C$* (at *179 K)*. But importantly, the ordered *$LaBaCo_2O_6$* depicts an *MR* value up to *14.5%* at *10 K* in an applied field of *±7 T* which is much larger than an *MR* value of *4%* at the same temperature (*10 K*) for the disordered phase. This difference suggests that at low temperature (*T < 50 K*), the grain boundary effect plays an important role in the anisotropic *MR* behavior for the ordered *$LaBaCo_2O_6$* perovskite. This is in agreement with its much larger upturn of resistivity, which is almost *10* times larger (at *10 K*) than the disordered *$La_{0.5}Ba_{0.5}CoO_3$* cobaltites. Thus, the larger *MR* observed for the ordered *$LaBaCo_2O_6$* is interpreted as tunnelling magnetoresistance (*TMR*) effect due to the increase of the intergrain insulating barriers [28], rather than an intrinsic effect, and is dominant at *10 K* over the *TMR* effect for the disordered phase. Therefore the spin-polarized tunnelling of carriers across the insulating boundaries occurring at the interfaces between polycrystalline grains give rise to the *TMR* effect in these phases. Figure 25, shows the *MR* effect for ordered *$LaBaCo_2O_{5.5}$* at five different temperatures. Unfortunately, in the case of temperature dependent resistivity data (with field), there is no major change in the *ρ(T)* curve even though the *M(T)* behavior reveals some kind of *FM* and *AFM* ordering. The charge transport for this kind of system is expected to be very sensitive due to the co-existence of *FM* and *AFM* state and the external magnetic fields readily induces an *MR* effect by affecting the subtle balance between *FM-AFM* phases. Unlike, disordered *$La_{0.5}Ba_{0.5}CoO_3$* and ordered *$LaBaCo_2O_6$* cobaltites, the highest negative *MR* value is obtained at *245 K* around *5 %* in an applied field of *7 T* and near room temperature the value is only *1.6 %*. At low temperatures, the *MR* values are only *2.8 %* (at *150 K*) and *2.5 %* (at *50 K*) respectively. The evidence of negative MR at low temperatures, similar to the *TMR* observed usually in polycrystalline samples, in the case of ordered cobaltites, is considered to be related to the spin-dependent scattering at grain boundaries. Nevertheless, in the case of ordered *$LaBaCo_2O_{5.5}$*, the highest *MR* value is noticed near the *FM-AFM* phase boundary, hence the grain boundary effect can be ignored and considered to be as an intrinsic effect.

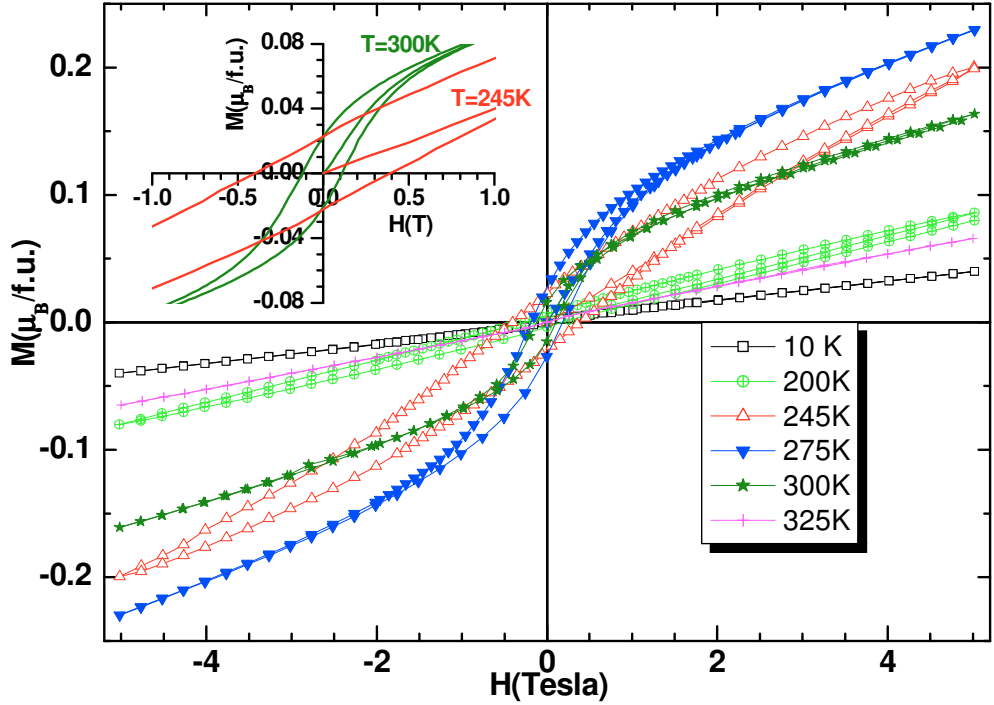

Figure 21. Magnetic field dependent isotherm magnetization, M(H), of ordered *LaBaCo$_2$O$_{5.5}$* at six different temperatures. The inset figure shows the expanded version for lower magnetic fields at 245 and 300K (Taken from Ref. 25).

Interestingly, the isothermal *MR* behavior at *245 K* exhibits an irreversible effect analogous to those of isothermal magnetization, *M(H)*, behavior (inset Figure 25), also present in *300 K* isothermal *MR* data. The peak in the isothermal *MR* data occurs around the coercive field value, which corresponds to the state of maximum disorder in the orientation of the neighbouring magnetic spins. Hence, the field dependent *MR* data that is indirectly related to the alignment between magnetic spins, reaches a maximum value. This effect is prominent for *300 K* data, compared to *245 K* as evidenced from the inset Figure 25 (dotted vertical lines), which may be due to *FM*-like state near *300 K* whereas latter one corresponds to magnetic phase boundary. Additionally, the isothermal *MR* data exhibit hysteresis effects that resemble the "butterfly-like" feature, although the effect is rather weak at low temperature (*50 K*). It is clear from the obtained data that the *MR* effect is strongly irreversible near the *FM-AFM* phase boundary. The "butterfly-like" feature appears only near the magnetic phase boundary. Correspondingly, the magnetic field dependent isotherm *MR* behavior at *10 K* for stoichiometry ordered-disordered phases exhibits an anisotropic effect similar to those of the magnetization behavior. Nevertheless, the isotherm *MR* data at low temperature for all three phases exhibit hysteresis effects, which resemble the "butterfly-like" feature, although the effect is rather weak for the disordered phase. It is noticed that the *MR* effect is almost isotropic for temperatures near or above the $T_C$. Hence, the butterfly-like feature appears only at low temperatures, which is prominent for ordered *LaBaCo$_2$O$_{5.5}$* and *LaBaCo$_2$O$_6$* phases. The origin of the magnetic field induced maximum *MR* near $T_C$ for the disordered $La_{0.5}Ba_{0.5}CoO_3$ and ordered $LaBaCo_2O_6$ are explained by the mechanism of suppression of spin fluctuations below $T_C$. On the other hand, the highest obtained *MR* value, obtained at *10 K* for the ordered $LaBaCo_2O_6$ is explained by the *TMR* effect due to the presence of more insulating grain boundaries. Hence, the occurrence of irreversible *MR* behavior nearly at similar temperatures as those for magnetic field variation isothermal *M(H)* studies suggest the strongly correlated nature of field-induced magnetic and electronic transitions [12, 25].

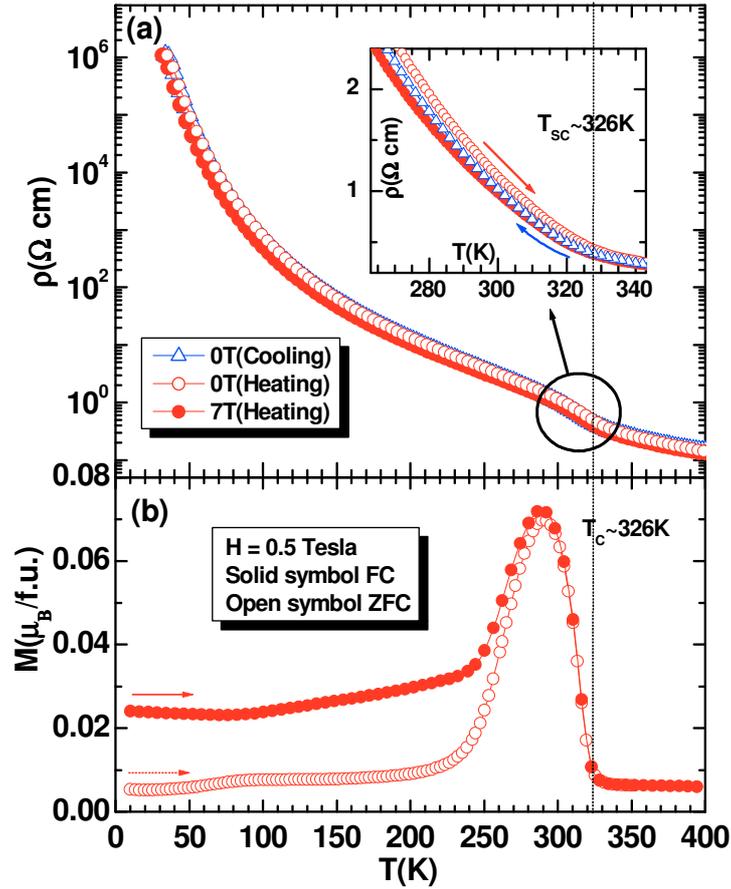

Figure 22. Temperature dependent physical properties for ordered $LaBaCo_2O_{5.5}$: (a) electrical resistivity, ρ(T), in the presence (solid symbol) and absence (open symbol) of magnetic field (7 Tesla) during heating and cooling cycles (inset shows the expanded version near the transition temperature, $T_{IM}$, and (b) ZFC (open symbol), FC (solid symbol) Magnetization in an applied field of 0.5 Tesla (Taken from Ref. 25).

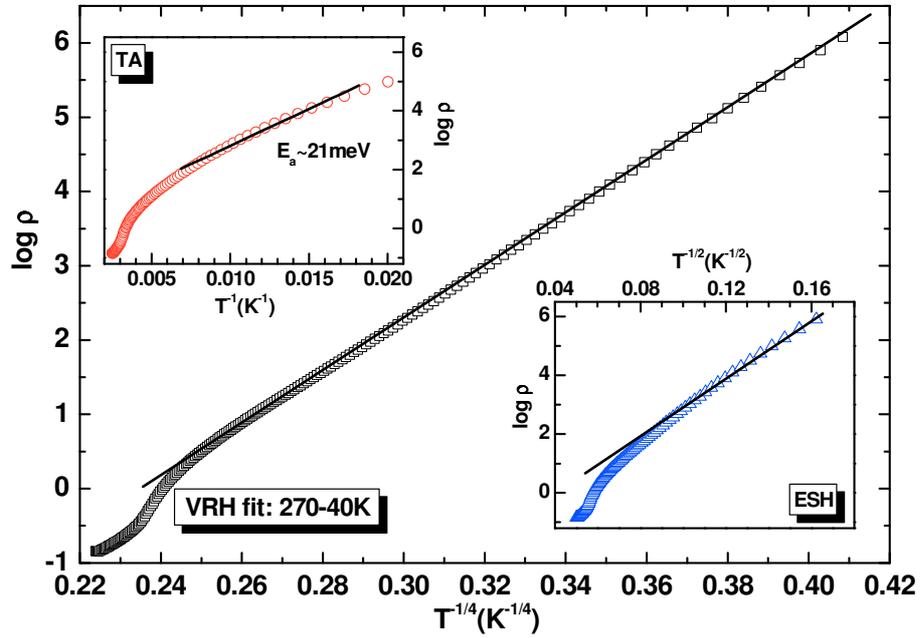

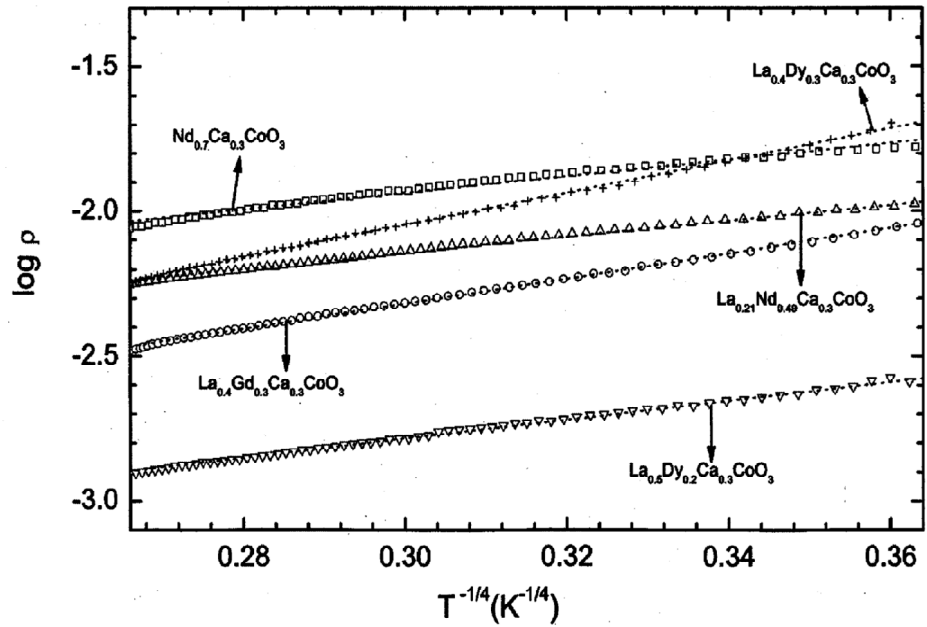

Figure 23. Logarithm of the resistivity versus $T^{-1/n}$ plots for (a) ordered $LaBaCo_2O_{5.5}$ and (b) disordered $Ln_{0.7}Ca_{0.3}CoO_3$ cobaltites (where n = 1, 2 or 4): open symbols and solid lines represent the experimental data and apparent fit to the hopping models as described in the text (Taken from Ref. 11 & 25).

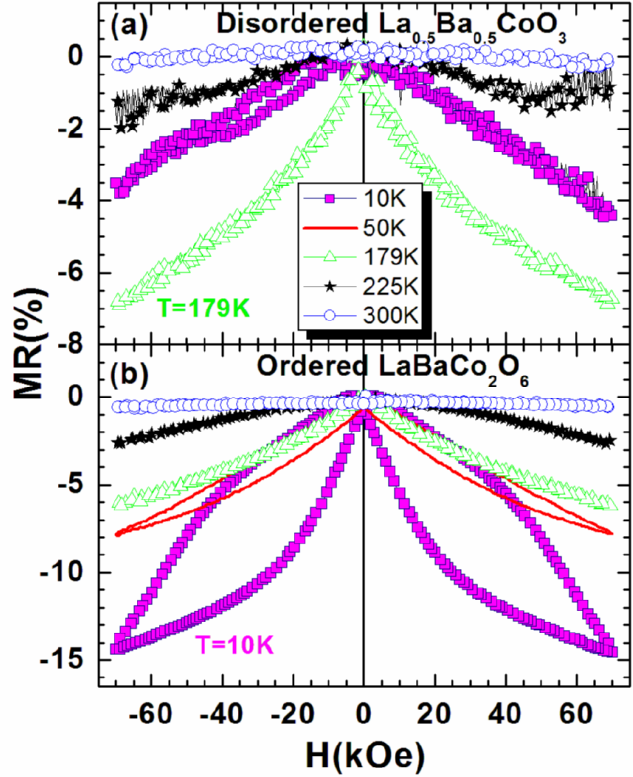

Figure 24. Temperature variation marnetoresistance for (a) disordered $La_{0.5}Ba_{0.5}CoO_3$ (b) ordered $LaBaCo_2O_6$ cobaltites (Taken from Ref. 12).

In order to get more insight into the nature of the conduction mechanism below $T_{SC}$, we have also discussed the thermopower (*Seebeck* coefficient), $S(T)$, and thermal conductivity, $\kappa(T)$, measurements for ordered $LaBaCo_2O_{5.5}$. These are sensitive to the magnetic and electrical nature of charge carriers (hole/electron) and can give some valuable information, which are absent in the magnetotransport measurements. Additionally, the $S(T)$ data is less affected due to the presence of grain boundaries, which often complicates the $\rho(T)$ data interpretation for polycrystalline samples. We noticed that, at room temperature the system shows a relatively large positive value of the thermoelectric power (*91 μV/K*), while it has a low resistivity (~*1.14 Ω cm* at *300 K*). The $S(T)$, for cooling and heating cycles is shown in Figure 26. It is observed that, the $S(T)$ value is positive below $T_{SC}$ and with decreasing temperature the value increases gradually and reaches a maximum value of ~*303 μV/K* at around *120 K* (marked by an arrow in the figure). Below that, the $S(T)$ value decreases continuously and similar to the literature behaviors (i.e. for $NdBaCo_2O_{5.5}$, $GdBaCo_2O_{5.5}$ and $HoBaCo_2O_{5.5}$ the peak values are *105, 88* and *70 K* respectively [26, 27]).

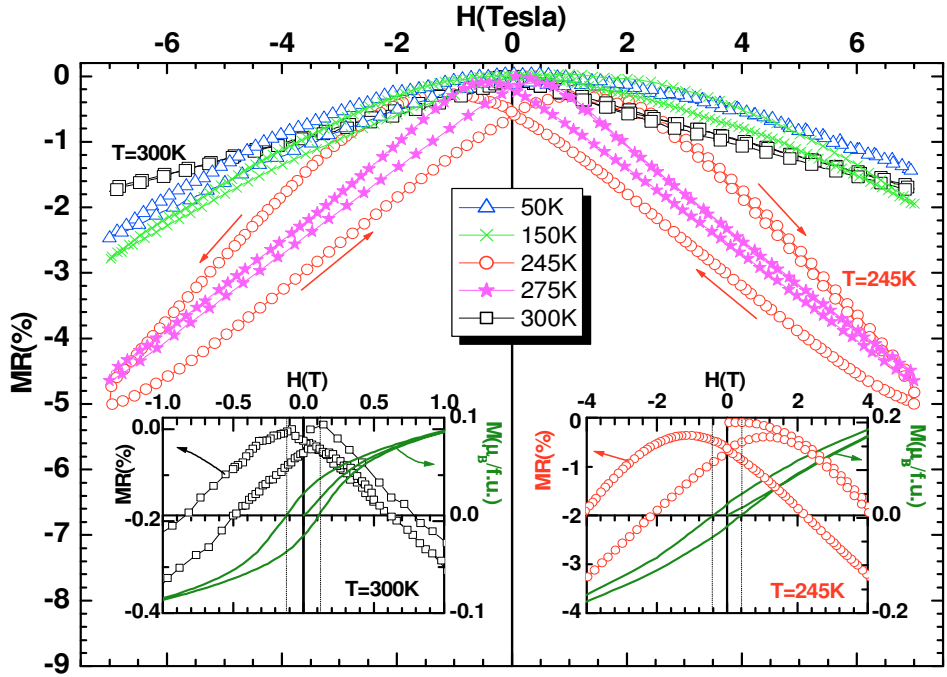

Figure 25. Magnetic field dependent isotherm magnetoresistance, MR, effect for ordered $LaBaCo_2O_{5.5}$ at five different temperatures (H= ±7 Tesla). Inset figures show the isotherm magnetization, M(H), and MR plot at 245 and 300 K for comparison (Taken from Ref. 25).

The $S(T)$ data implies that the $LnBaCo_2O_{5.5}$ systems show a semiconducting type behavior of the thermopower down to low temperature akin to resistivity behavior and below the $S(T)$ behavior is complex, decreasing abruptly at low temperature. In fact, with decreasing temperature the $S(T)$ value should increase due to trapping or localization of charge carriers. This type of $S(T)$ behavior is quite unexpected for semiconducting thermoelectric materials and there is no general explanation till date. Following the general approach in analyzing the semiconducting behavior we have plotted the $S(T)$ data in the $T^{-1/n}$ scale similar to $\rho(T)$. Since for semiconductors the $S(T)$ is expected to be linear in $T^1$ behavior (according to TA model) or to follow either of the described hopping models similar to resistivity behavior [11, 25]. It is noticed that the $S(T)$ data can not be described by the previous mentioned models and one observes p-type conductivity throughout the temperature range. Many authors have already explained the $S(T)$ behavior at higher temperature and their sign reversal near the electronic transition [26, 27]. However, the low temperature $S(T)$ evolution, basically the appearance of the broad maximum in the metastable AFM phase and the decreasing nature with temperature (in spite of semiconducting behavior) has not been explained properly. We have analysed the obtained thermopower data using an expression $S(T) = S_0 + S_{3/2}T^{3/2} + S_4T^4$ defined by *P. Mandal* [29], which can be understood on the basis of electron magnon scattering (spin wave theory). In ferro- and antiferromagnets, electrons are scattered by spin wave as explained earlier for perovskite manganites [29] and hence it is expected that this theory may explain the $S(T)$ behavior for the present system. However, the obtained thermopower value is higher than the one reported for manganites. The $S(T)$ data in *60 K ≤ T ≤ 105 K* range follows $T^{3/2}$ behavior and in the *120-320 K* range, it fits linearly with the $T^4$ behavior (insets of Figure 26).

At low temperature the second term ($S_{3/2}$) will dominate over $S_4$ ($S_{3/2} \gg S_4$), hence the *S(T)* will depict downward trend at low temperature. Although we do not have sufficient experimental data points below the broad peak, yet the *S(T)* curve fits linearly to the $T^{3/2}$ term (inset of Figure 26b), as expected from the spin wave theory. Additionally, the downward feature in *S(T)* data is present in all studied $LnBaCo_2O_{5.5}$ cobaltites [25-27]. Hence, the broad peak at low temperature and downward trends for layered cobaltites are considered to be the result of the electron magnon scattering akin to perovskite manganite as explained by *P. Mandal* [29].

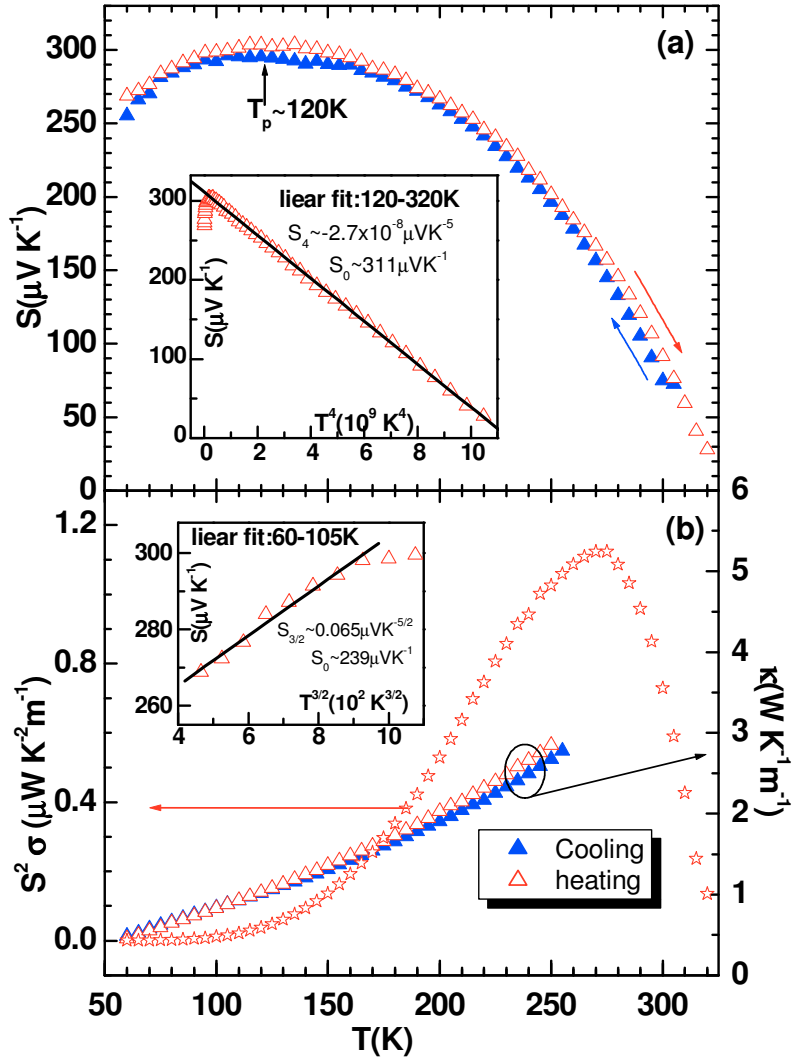

Figure 26. Temperature dependent transport measurements for ordered $LaBaCo_2O_{5.5}$: (a) thermoelectric power, S(T), during cooling (solid triangle) and heating (open triangle) cycles and inset shows the S-$T^4$ plot in the 120-320K range, (b) thermal conductivity, κ(T), and power factor, $S^2\sigma$(T), in the temperature range of 60-320K (inset shows the S-$T^{3/2}$ plot in the 60-105K range) (Taken from Ref. 25).

Thermal conductivity, $\kappa(T)$, measurements of $LaBaCo_2O_{5.5}$ during cooling and heating cycles are shown in Figure 26b. The $\kappa(T)$ value increases with increasing temperature, indicating an usual phonon mediated scattering mechanism of charge carriers, which is the sum of phonon ($\kappa_p$) and electronic ($\kappa_e$) contribution [*i.e.* $\kappa(T) = \kappa_p(T) + \kappa_e(T)$]. Therefore, from the *Wiedemann-Franz* law we have calculated the electronic contribution ($\kappa_e$) at *250 K*, which is around ~$1.5 \times 10^{-4}$ $WK^{-1}m^{-1}$ [whereas $\kappa(250\ K) \sim 2.8\ WK^{-1}m^{-1}$]. This implies that, the lattice modulated phonon contribution ($\kappa_p$) is dominant near room temperature *i. e.* $\kappa_p(T) >> \kappa_e(T)$. We have also calculated the temperature dependence of power factor ($S^2\sigma$) from the obtained resistivity and thermopower data as shown in Figure 26b, which depicts a clear maximum near *270 K* and decreases rapidly on both sides of the temperature scale (*60K ≤ T ≤ 320K*). It is important to note that, this maximum is in the region of *FM-AFM* phase boundary. This suggests that the electron magnon scattering at low temperature strongly affects the *S(T)* behavior of $LaBaCo_2O_{5.5}$ system. The calculated figure of merit (*ZT*) shows too small values close to room temperature (~$10^{-5}$) for applications as expected from 112-type cobaltite systems.

## CONCLUSION

The discussion in the earlier sections clearly brings out the varied effects of magnetization, magneto-transport properties, electronic phase separation, thermopower etc, in the ordered-disordered rare earth cobaltites. This paper shows that the method of synthesis – precursor reactivity and oxygen pressure – plays a crucial role for controlling the cationic order-disorder phenomena in the non-stoichiometry ordered $LaBaCo_2O_{5.5}$ and the stoichiometric disordered $La_{0.5}Ba_{0.5}CoO_3$ and ordered $LaBaCo_2O_6$. From the viewpoint of the transport properties, the ordered $LaBaCo_2O_{5.5}$ is semiconducting throughout and the disordered $La_{0.5}Ba_{0.5}CoO_3$ exhibits a semi-metal to metal transition around $T_C$, whereas the layer-ordered $LaBaCo_2O_6$ is characterized by semi-conductor to metal transition around $T_C$. As a consequence, they exhibit a similar intrinsic magnetoresistance, maximum at the vicinity of $T_C$ (*6 to 7 %*), whereas the ordered $LaBaCo_2O_6$ exhibits a much larger *MR* value, indicating *TMR* at low temperature due to different grain boundary effects. It appears that the phenomenon is much more common than anticipated, to the extent that some workers suggest that even the physical properties of these cobaltites, showing different types of interesting behavior, is a consequence of cationic ordering. It is noteworthy that the large thermopower value caused by lowering temperature in ordered $LaBaCo_2O_{5.5}$ may indeed be due to semiconducting behavior. The unusual *S(T)* behavior and the appearance of a broad peak at low temperature with a maximum value of ~*303 µV/K* is explained by electron magnon scattering mechanism. This is expected to be applicable to all series of $LnBaCo_2O_{5.5}$ cobaltites at low temperature. It would be worthwhile to study the relevance of this phenomenon to other transition metal oxide systems.

## ACKNOWLEDGEMENTS
AKK gratefully acknowledges Prof. A. Ojha, Director of IIITDM-J, for sanctioning faculty research grant.